\documentclass[sigconf]{acmart}

\usepackage{tikz}
\usepackage{fancyhdr}
\usepackage{amsfonts, amsmath}
\usepackage{booktabs}
\usepackage{multirow}
\usepackage{graphicx, xcolor, setspace, url, verbatim, afterpage}
\usepackage{stfloats}
\usepackage{xspace}
\usepackage{paralist}
\usepackage{textcomp}
\usepackage{algorithm}
\usepackage[noend]{algpseudocode}
\usepackage[formats]{listings}
\usepackage{caption}

\usepackage{soul,color}

\usepackage[strict]{changepage}

\usepackage{framed}

\definecolor{reccshade}{rgb}{0.95,0.95,1}
\definecolor{warshade}{rgb}{1,0.95,0.85}
\definecolor{greenshade}{rgb}{0.95,1,0.95}
\definecolor{darkgreen}{rgb}{0,0.5,0}
\definecolor{darkblue}{rgb}{0,0,0.7}

\newenvironment{recommendation}{%
	\MakeFramed{\advance\hsize-\width\FrameRestore}%
	\noindent\hspace{-4.55pt}
	\begin{adjustwidth}{}{7pt}%
	\vspace{2pt}\vspace{2pt}%
}
{%
		\vspace{2pt}\end{adjustwidth}\endMakeFramed%
}

\newenvironment{warning}{%
	\MakeFramed{\advance\hsize-\width\FrameRestore}%
	\noindent\hspace{-4.55pt}
	\begin{adjustwidth}{}{7pt}%
		\vspace{2pt}\vspace{2pt}%
	}
	{%
		\vspace{2pt}\end{adjustwidth}\endMakeFramed%
}

\newenvironment{take}{%
	\MakeFramed{\advance\hsize-\width\FrameRestore}%
	\noindent\hspace{-4.55pt}
	\begin{adjustwidth}{}{7pt}%
		\vspace{2pt}\vspace{2pt}%
	}
	{%
		\vspace{2pt}\end{adjustwidth}\endMakeFramed%
}

\newcommand{\secref}[1]{Section~\ref{#1}}
\newcommand{\secsref}[2]{Sections~\ref{#1} and~\ref{#2}}
\newcommand{\figref}[1]{Figure~\ref{#1}}
\newcommand{\figsref}[2]{Figures~\ref{#1} and~\ref{#2}}

\newcommand{\tabref}[1]{Table~\ref{#1}}
\newcommand{\ignore}[1]{}
\newcommand{\minititle}[1]{\textbf{#1}}

\def\everywhere{\textsf{MPI everywhere}}
\def\threads{\textsf{MPI+threads}}
\def\shared{\textsf{MPI shared memory}}

\def\multiple{\texttt{MPI\_THREAD\_MULTIPLE}}
\def\funneled{\texttt{MPI\_THREAD\_FUNNELED}}

\def\mpiinit{\texttt{MPI\_Init}}
\def\mpifinalize{\texttt{MPI\_Finalize}}

\def\mpiwinfree{\texttt{MPI\_Win\_free}}

\def\mpisend{\texttt{MPI\_Send}}
\def\mpissend{\texttt{MPI\_Ssend}}
\def\mpiisend{\texttt{MPI\_Isend}}
\def\mpiirecv{\texttt{MPI\_Irecv}}

\def\mpiput{\texttt{MPI\_Put}}
\def\mpiget{\texttt{MPI\_Get}}
\def\mpiaccum{\texttt{MPI\_Accumulate}}
\def\mpifetchandop{\texttt{MPI\_Fetch\_and\_op}}

\def\mpiflush{\texttt{MPI\_Win\_flush}}

\def\mpiwait{\texttt{MPI\_Wait}}

\def\anysource{\texttt{MPI\_ANY\_SOURCE}}

\def\netmod{\texttt{netmod}}
\def\shmmod{\texttt{shmmod}}
\def\core{\texttt{ch4\_core}}

\def\naive{na\"\i ve}

\def\ie{{\it i.e.}}
\def\eg{{e.g.}}

\AtBeginDocument{%
  \providecommand\BibTeX{{%
    \normalfont B\kern-0.5em{\scshape i\kern-0.25em b}\kern-0.8em\TeX}}}

\setcopyright{rightsretained}

\begin{document}
	
\copyrightyear{2020}
\acmYear{2020}
\acmConference[ICS '20]{2020 International Conference on Supercomputing}{June 29-July 2, 2020}{Barcelona, Spain}
\acmBooktitle{2020 International Conference on Supercomputing (ICS '20), June 29-July 2, 2020, Barcelona, Spain}\acmDOI{10.1145/3392717.3392773}
\acmISBN{978-1-4503-7983-0/20/06}

\title{How I Learned to Stop Worrying about\\User-Visible Endpoints and
  Love MPI}

\author{Rohit Zambre}
\email{rzambre@uci.edu}
\affiliation{
	\institution{University of California, Irvine}
}

\author{Aparna Chandramowliswharan}
\email{amowli@uci.edu}
\affiliation{
	\institution{University of California, Irvine}
}

\author{Pavan Balaji}
\email{balaji@anl.gov}
\affiliation{
	\institution{Argonne National Laboratory}
}

\renewcommand{\shortauthors}{Zambre et al.}

\begin{abstract}
	
MPI+threads is gaining prominence as an alternative to the traditional
``MPI everywhere'' model in order to better handle the
disproportionate increase in the number of cores compared with other
on-node resources.  However, the communication performance of
MPI+threads can be 100x slower than that of MPI everywhere.  Both MPI
users and developers are to blame for this slowdown.  MPI users
traditionally have not exposed logical communication parallelism.
Consequently, MPI libraries have used conservative approaches, such as
a global critical section, to maintain MPI's ordering constraints for
MPI+threads, thus serializing access to the underlying parallel network resources
and limiting performance.

To enhance the communication performance of MPI+threads, researchers
have proposed MPI Endpoints as a user-visible extension to the MPI-3.1
standard.  MPI Endpoints allows a single process to create multiple
MPI ranks within a communicator.  This could, in theory, allow each
thread to have a dedicated communication path to the network, thus
avoiding resource contention between threads and improving
performance.  The onus of mapping threads to endpoints, however, would
then be on domain scientists.  In this paper we play the role of
devil's advocate and question the need for such user-visible
endpoints.  We certainly agree that dedicated communication channels
are critical.  To what extent, however, can we hide these channels
inside the MPI library without modifying the MPI standard and thus
unburden the user?  More important, what functionality would we lose
through such abstraction?  This paper answers these questions through
a new implementation of the MPI-3.1 standard that uses multiple
virtual communication interfaces (VCIs) inside the MPI library.  VCIs
abstract underlying network contexts.  When users expose parallelism
through existing MPI mechanisms, the MPI library maps that parallelism
to the VCIs, relieving the domain scientists from worrying about
endpoints. We identify cases where user-exposed parallelism on VCIs
perform as well as user-visible endpoints, as well as cases where such
abstraction hurts performance.

\end{abstract}




\maketitle

\section{Introduction}
\label{sec:introduction}

\everywhere\ (typically one MPI process per core) has been the
traditional model for using MPI on supercomputers.  While the model
has served applications well for several decades, it is becoming
difficult to scale on modern architectures, primarily owing to the
disproportionate increase in the number of cores per node compared
with other on-node resources such as memory and network registers.
For example, memory wastage for halo regions in PDE simulations with
the \everywhere\ model is a known
problem~\cite{rabenseifner2009hybrid}, which worsens with the increase
in dimensionality of the domain decomposition.  To address this issue,
researchers have been increasingly adopting hybrid
\threads\ (typically one MPI process per node or socket, and one
thread per core) parallelism (\eg,
MPI+OpenMP)~\cite{bernholdt2017survey} since it allows them to utilize
the many cores on a node while sharing the remaining on-node
resources~\cite{jin2011high,bulucc2017distributed,higgins2015hybrid}.

The communication performance of \threads{}, however, is dismal,
especially when multiple threads are involved in communication (i.e.,
\multiple).  The reason for the poor performance stems from the quaint
view, held by both MPI users and MPI developers, of the network as a
sequential hardware resource.  Modern network interface cards (NICs)
feature multiple hardware communication contexts that allow for
independent, parallel communication streams from a single
node~\cite{zambre2018scalable}.  To efficiently utilize the network
parallelism, MPI users must expose logical parallelism in their
communication so that threads can map to different underlying hardware
contexts on the NIC.  How does one expose such logical communication
parallelism?  The MPI standard specifies certain sequential ordering
constraints between messages~\cite{nonovertakingorder} that guarantee
some determinism in execution.  But these sequential ordering
constraints are based on $<$communicator, rank, tag$>$ or $<$window,
rank$>$ tuples.  The user can inform the MPI library that two or more
messages have no relative ordering between them by using, for example,
different communicators, different windows, or in some cases different
ranks or tags, thus exposing logical parallelism between these
messages.

The state of the art, however, is conservative in this
regard. Applications typically do not expose MPI communication
parallelism because MPI libraries today do not utilize such
parallelism. MPI libraries, on the other hand, still employ
conservative approaches, such as a global critical section and the use
of only one network hardware context, because applications today do
not expose any parallelism that the library can exploit.

To facilitate the improvement in the communication performance of
\threads, user-visible MPI
Endpoints~\cite{mpiendpointsproposal,dinan2013enabling} has been
proposed as an extension to the MPI-3.1 standard. These user-visible
endpoints allow the user to explicitly map threads to the underlying
network resources.  If the user mapped each thread to a distinct
endpoint, then, in theory, all threads could have dedicated
communication paths to the network.  Several
efforts~\cite{sridharan2014enabling,dinan2014enabling,holmesintroducing}
indeed demonstrate scaling communication throughput with MPI Endpoints
through dedicated communication channels inside the MPI library.
Unfortunately, these works have a number of shortcomings. First and
foremost, user-visible solutions  enforce the new concept of endpoints upon
users through new APIs and put the onus of mapping threads to the
endpoints on the domain scientist, potentially hurting productivity.
Second, prior works did not compare against an MPI-3.1 implementation that uses
multiple network hardware contexts, thus implicitly assuming that current
implementations of MPI-3.1 are already the most optimal. Third, they
modify applications to expose communication parallelism to the MPI
library with MPI Endpoints but do not expose the equivalent
parallelism to the MPI-3.1 version of the library, thus making the
comparison unfair.  Finally, irrespective of whether we optimize the
implementation of MPI-3.1 or implement MPI Endpoints, certain corner
cases with respect to communication progress must be handled for
correctness, even though they sometimes hurt performance.  Prior works
ignore such corner cases, sacrificing correctness in pursuit of higher
performance.

In this paper, we play the role of devil's advocate to user-visible
endpoints.  In fairness to the previous efforts, we agree with them on
two aspects: (1) applications must expose communication parallelism,
but MPI-3.1 already provides multiple mechanisms to do that; and (2)
MPI libraries must provide multiple independent communication
channels, but an efficient MPI library can do so internally without
exposing them to users as endpoints.  Thus we are left with the
question: Are extensions to the MPI standard necessary in order to
improve \threads\ communication?

To answer this question, we start with a new MPI-3.1 library that
internally uses multiple virtual communication interfaces (VCIs).  A
VCI represents a communication stream that is mapped to a network
hardware context.  Users expose communication parallelism through
existing MPI mechanisms (such as communicators, windows, ranks, and
tags), and the MPI library maps that parallelism to the different
hardware contexts by funneling messages over the internal VCIs.  More
important, VCIs are completely hidden within the MPI library, thus
requiring no extension to the MPI standard and placing no requirement
for thread-to-network-resource mapping on the domain scientists.

The effectiveness of transparently using multiple VCIs depends on the
communication pattern of applications.  In this regard, we classify
\threads\ applications into three categories: (1) applications that
can directly use dedicated communication channels where the multi-VCI
approach saturates the network performance similarly to \everywhere{}
and user-visible endpoints; (2) applications that require
shared progress where both multiple VCIs and user-visible endpoints
suffer from loss in performance; and (3) applications that need direct
access to the network resources where abstracting the VCIs can hurt
performance compared with user-visible endpoints.  We study
applications in all three categories in this paper.  To that end, we
make the following contributions:

\begin{enumerate}

\item We develop a fast \threads\ library by addressing thread safety
  and network resource underutilization while adhering to the MPI
  standard (in \secref{sec:mpi_library_opts}).
	
\item We compare the capabilities of MPI-3.1 with those of
  user-visible endpoints for microbenchmarks and real applications (in
  \secsref{sec:microbench_eval}{sec:miniapps}).
	
\item We provide users with recommendations on exposing communication
  parallelism in their applications with MPI-3.1 (in
  \secref{sec:miniapps}) based on \secref{sec:mpi_parallelism}'s discussion
  of parallelism in the existing MPI standard.

\end{enumerate}

\section{Parallelism in the MPI Standard}
\label{sec:mpi_parallelism}

For both two- and one-sided MPI communication, the existing MPI standard
allows applications to expose communication parallelism.  Below, we
discuss approaches for a single MPI process to expose such
parallelism.  In particular, communication parallelism in MPI can be
viewed as multiple independent communication streams, where each
stream is a first-in, first-out (FIFO) ordered set of communication
operations.

\subsection{Point-to-point communication}
\label{sec:pt2tp_parallelism}

For two-sided communication, MPI uses the <communicator, rank, tag>
triplet to match operations.

\minititle{Different communicators.} MPI does not define any order
between operations executed on different communicators.  This approach
implies that all operations on different communicators can execute
independently on parallel communication streams.

\begin{figure*}[htbp]
	\begin{center}
		\includegraphics[width=0.99\textwidth]{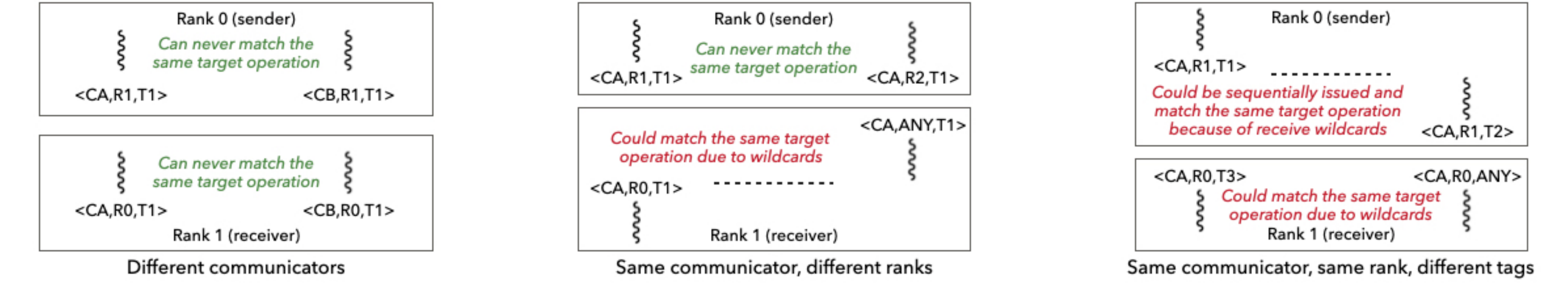}
	\end{center}
	\caption{Different combinations of <comm,rank,tag> tuples demonstrating
		point-to-point parallelism in the MPI standard. Dashed horizontal lines
	represent thread barriers.}
	\label{fig:pt2pt_parallel_ex}
\end{figure*}

\minititle{Same communicator, different ranks.}  Within a
communicator, MPI specifies a \emph{nonovertaking
  order}~\cite{nonovertakingorder}: if multiple ordered operations
match the same target operation, the operation that was issued first
must consume the target operation before the one that was issued
later.  No matching order applies to operations intended for different
targets.  For example, no ordering constraints apply to multiple send
operations that use the same communicator but target different ranks.
Hence, they can execute on parallel communication streams.  On the
other hand, receive operations that use the same communicator
cannot execute in parallel  even if they specify different ranks.
The reason is that it is possible for any receive operation to contain
the \anysource\ wildcard.  To ensure correct matching order, the
MPI library needs to funnel all receive operations of a communicator
through the same communication stream (see \figref{fig:pt2pt_parallel_ex}).

\minititle{Same communicator, same rank, different tags.}
Operations that target the same rank within a communicator but
use different tags cannot utilize parallel communication streams for
both send and receive operations. The order of operations in
MPI is determined by the MPI user.  In \threads{}, operations on
different threads may be parallel or ordered through, for example, a
thread barrier.  Suppose the user issues two operations on two
different threads with a barrier between the operations  (see
\figref{fig:pt2pt_parallel_ex}).  A target
operation that satisfies both operations must first match the
operation that was issued before the barrier.  To ensure this, the MPI
library must use the same communication stream for the operations from
the two threads.  If the operations use different communication
streams, the operation issued after the barrier could incorrectly
match the target prior to the one issued before the barrier.

\subsection{Remote Memory Access communication}
\label{sec:rma_parallelism}

MPI's one-sided communication, namely, remote memory access (RMA), is
executed on top of windows. Unlike point-to-point, RMA operations do
not have any matching constraints and feature a lot more parallelism. MPI
does not require any ordering for its Get, Put, and Accumulate classes of
operations if two or more operations target different ranks or use different
windows. Additionally, two or more Put or Get operations do not have
any ordering constraints even if they use the same window. Hence, multiple
Get and Put types of operations can execute on parallel communication
streams. But, by default, MPI-3.1 requires program order for Accumulate 
operations originating from the same source and targeting the same memory
location on the same window. It does, however, give the user the option
to relax this ordering constraint through the \texttt{accumulate\_ordering}
hint. Without hints, multiple Accumulate-style operations
can execute on parallel communication streams if they use different
windows or target different memory locations.

Even though multiple RMA operations on the same window could use
parallel communication streams, mixing synchronization operations,
such as \mpiflush{}, with communication operations, such as \mpiget{},
can be tricky. Synchronization calls can wait for both past and concurrent
communication operations to complete. Thus, if one thread is waiting
inside \mpiflush{} and another thread continuously issues \mpiget{}
operations, the first thread might block indefinitely. Apart from these
constraints, all types of RMA operations on different windows can execute
through separate communication streams in parallel.
\section{Software and testbeds}
\label{sec:testbed}

Our MPI implementation is based on the highly optimized
CH4~\cite{raffenetti2017mpi} device of the MPICH library.  The CH4
device is a combination of three components: a core (\core), a network
module (\netmod), and a shared-memory module (\shmmod).  The
\netmod\ and \shmmod\ are responsible for conducting internode and
intranode communication, respectively.  In this work, we focus on the
\netmod\ component because we assume \threads\ applications would
directly use the shared memory of the process for intranode
communication.

For most common data operations, CH4 offloads functionalities, such as
tag matching, to the low-level communication library, such as
OpenFabrics Interfaces (OFI)~\cite{grun2015brief} or Unified
Communication X (UCX)~\cite{shamis2015ucx}.  Where the hardware cannot
independently handle operations, CH4 falls back on using an active
message implementation of the operation in its \core.

Our testbeds include two platforms: the Skylake cluster and the Gomez
cluster in the Joint Laboratory for System Evaluation at Argonne
National Laboratory.  The clusters feature different interconnects:
Skylake hosts Intel Omni-Path (OPA) and Gomez hosts Mellanox
InfiniBand (IB) EDR.  These two families of interconnects constitute
the majority of the TOP500 in the supercomputing
space~\cite{top500interconnects}.  For Skylake, we use the OFI
\netmod\ in conjunction with PSM2; for Gomez, we use the UCX
\netmod\ with Verbs.

For our analysis and evaluation, we use the cores on the socket that
the NIC is attached to.  We ensure that the CPU speed is set to its
base frequency and that turbo boost is turned off.

\section{A Fast MPI+Threads Library}
\label{sec:mpi_library_opts}

In this section, we detail our implementation of parallel
communication streams (or VCIs) within a single MPI process.  Although
our work is on MPICH, the concepts extend to other MPI libraries as
well.  To design a fast \threads\ library, we need to deserialize
access to the software as well as to the network hardware resources;
the former is a critical precursor for the latter to extract
performance.

\subsection{Deserializing access to the MPI library}
\label{sec:fg_cs}

In \threads, the MPI library needs to protect its resources from the
threads' parallel updates.  State-of-the-art MPI implementations
conservatively employ a large global critical section with a single
lock.  The MPI operation enters the critical section at the beginning
of its execution and exits it either when it returns from the function
or when it yields to other threads to make progress.  This approach
largely serializes communication from multiple threads even if the
communication operations issued by those threads are independent.

\minititle{Fine-grained critical sections.} Balaji et
al.~\cite{balaji2008toward,balaji2010fine} and Amer et
al.~\cite{amer2019software} split the global lock in MPICH into
multiple locks such that each lock protects a different class of
objects.  For example, access to the network communication portal is
protected by a lock different from the one that protects the
management of request objects. Although fine-grained critical sections
(FG) mean higher parallelism, they incur two expenses over a global
critical section (Global): (1) more lock acquisitions and releases on
the critical path and (2) atomics for reference and completion
counters.

The number of locks taken in FG depends on the type of operations. For
any initiation operation, we need at least one lock---the one that
protects access to the communication portal.  Generally, for
\mpiisend\ and \mpiirecv , we need a second lock---the one that
allocates a request object from the global pool of requests.  For
small-message transmissions, however, we do not need the lock of the
request pool. Up to a certain message size, modern interconnects
guarantee completion as soon as they are posted; they do not require
any polling of the network.\footnote{A correct MPI implementation would
  need to poll the network intermittently even for such operations to
  progress any active message execution of an operation.} MPICH
optimizes memory usage for such operations by maintaining a global
lightweight request that is marked as complete. These operations then
simply increase the reference counter of the pre-completed request.

\begin{figure}[t!]
	\centering
	\begin{minipage}[t]{0.236\textwidth}
		\centering
		\includegraphics[width=\textwidth]{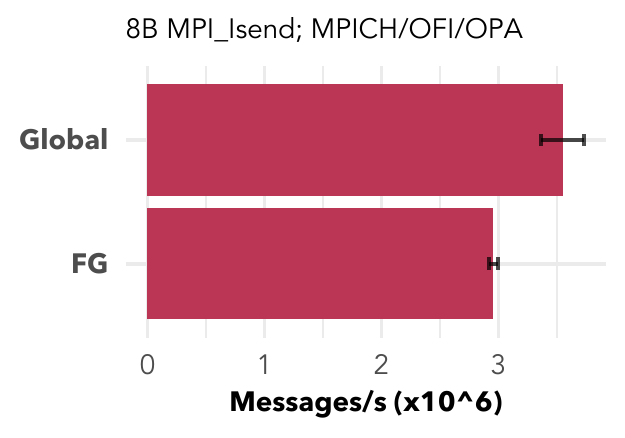}
		\caption{Overhead of FG.}
		\label{fig:thread_safety_opts}
	\end{minipage}
	\begin{minipage}[t]{0.236\textwidth}
		\centering
		\includegraphics[width=\textwidth]{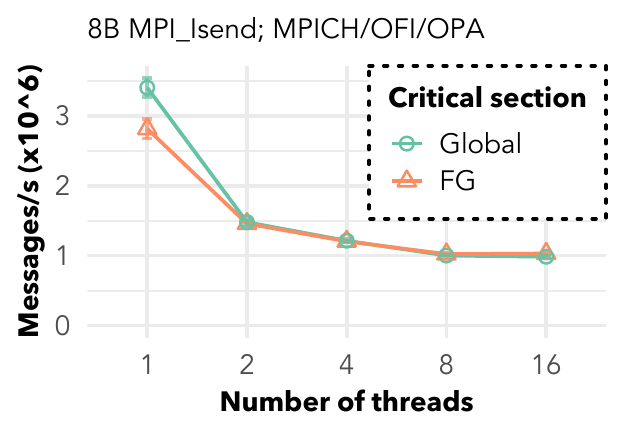}
		\caption{Global vs. FG.}
		\label{fig:global_vs_fg_scaling}
	\end{minipage}
\end{figure}

For progress operations, the number of locks taken depends on the
number of times the progress engine is invoked.  The progress engine not
only checks for the completion of an operation but also progresses
active outstanding schedules, such as those of non-blocking
collectives.  One iteration of the progress engine in MPICH takes
three locks: one to poll the communication portal and two
to check the activeness of progress hooks\footnote{MPICH/CH4 currently
  maintains two progress hooks.}  (each hook maintains its own thread
safety). When an operation completes, another lock is taken when the
request object is returned to the pool.

Although FG improves concurrency when multiple threads compete for MPI
resources, it adds some overhead when there is no contention (e.g.,
when a single thread is active).  \figref{fig:thread_safety_opts}
shows that FG hurts performance by 16.71\% in the uncontended case
(compared with Global).  This performance difference is due to the
higher number of locks and to atomic counting (as we corroborate in
\secref{sec:microbench_eval}).  With increasing number of threads, the
performance difference between FG and Global reduces, and FG
eventually outperforms Global at 16 threads, as seen in
\figref{fig:global_vs_fg_scaling}.  Moreover, although Global performs
better than FG for fewer threads, FG is critical when parallel
communication streams exist, as we show in
Section~\ref{sec:multi_vci_opts}.

\subsection{Parallel communication streams}
\label{sec:multi_vcis}

To address the problem of network resource underutilization, we first
define the virtual communication interface object. A VCI is an
abstract representation of a communication stream. Each VCI maps to a
communication context on the network hardware and contains its own
independent set of communication resources that maintain a FIFO order
of the MPI operations that map to it. Hence, with multiple VCIs, we get
parallel communication streams in the MPI library.  The physical
realization of a VCI depends on the \netmod\ and the underlying
interconnect. A VCI in the OFI \netmod\ is an OFI endpoint (for
transmission and reception) that is bound to an OFI completion queue
(for progress). For Intel OPA, the OFI endpoint maps to a hardware
context on the Intel HFI network adapter~\cite{hfi_guide}. A VCI in
the UCX \netmod\ is a UCP worker. For Mellanox IB, the UCP worker
contains Verbs resources: a queue pair (QP) for transmission, a shared
receive queue for reception, and a completion queue for progress. The
QP maps to the micro UARs (hardware registers) on the Mellanox
adapter~\cite{mlxRPM}.

\minititle{VCI pool design.}  To utilize the underlying network
parallelism, we maintain a pool of VCIs inside a single MPI process.
Since operations on different communicators can execute on different
communication streams (see \secref{sec:mpi_parallelism}), every time
the user creates a new communicator, we assign it a VCI from the pool
and mark that VCI as active. All operations on the communicator now
funnel through the VCI that was assigned to the communicator.  If
multiple threads communicated using separate communicators, they
would, in theory, establish parallel communication streams to the NIC
from the same process.  However, since the number of contexts on the
network hardware is limited,\footnote{For example, Intel OPA features
  only 160 hardware contexts on the HFI adapter~\cite{hfi_guide}} the
VCI pool may be empty during communicator creation. In such a case, we
revert to a fallback VCI. For this work, we designate the VCI
allocated to \texttt{MPI\_COMM\_WORLD} as the fallback.  When the user
frees a communicator, its associated VCI is returned to the pool and
is marked as inactive. Certainly better techniques to map
communicators to VCIs exist, but such techniques are out of the scope
of this paper and will be analyzed in the future.  The overhead from
this design is that each operation now needs to compute which VCI to
use on the critical path.  For the communicator-to-VCI mapping, this
computation is a lookup, which costs 8 additional instructions in our
implementation.  The VCI pool design extends to RMA operations as
well, where we assign VCIs to each window since operations on
different windows can execute in parallel (see
\secref{sec:mpi_parallelism}).

\minititle{Thread safety.} We extend the fine-grained critical
sections from \secref{sec:fg_cs} such that each VCI is then protected
by its own separate lock since it is independent. Threads that map to
different VCIs can access the VCIs without contention.

\begin{figure}[t!]
	\begin{center}
		\includegraphics[width=0.49\textwidth]{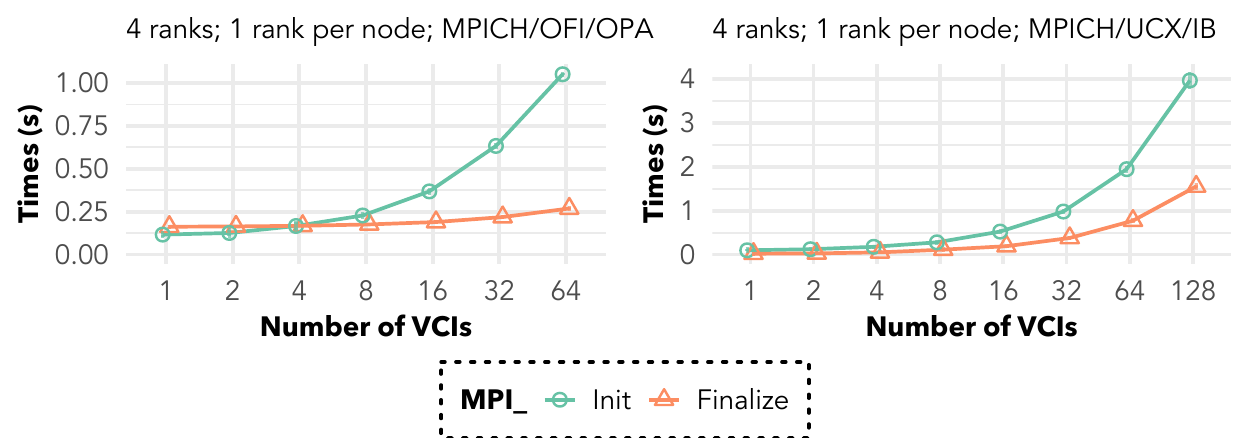}
	\end{center}
	\caption{Multi-VCI \mpiinit\ and \mpifinalize\ overheads.}
	\label{fig:init_finalize_time}
	\vspace{-1em}
\end{figure}

\minititle{Connection establishment.} Each VCI has its own
transport-level address that needs to be exchanged between the ranks
in order to establish connections. We do so during the initialization
of MPI. We first use PMI~\cite{balaji2010pmi} to exchange the
addresses of the fallback VCIs on every rank.  Using the fallback VCI,
we exchange the addresses of the rest of the VCIs using an allgather
operation. As expected, establishing connections statically during
initialization incurs an overhead that grows with the number of VCIs
(see \figref{fig:init_finalize_time}). Similarly, the finalization
time increases since the tear-down time of VCIs is proportional to the
number of VCIs.\footnote{Features like OFI scalable endpoints can
  reduce the connection establishment and tear-down overheads, because
  they share the same transport-level address.  However, we have not
  used them in this work because their performance is still not on par
  with that of regular endpoints, at least for the PSM2 provider that
  we used in this work.  Furthermore, scalable endpoints share some
  resources, such as the OFI address vector, accesses to which could
  be serialized in the critical path by the OFI
  provider~\cite{avsharing}.}


\begin{figure*}[htbp]
  \centering
  \begin{minipage}[t]{0.24\textwidth}
    \centering
    \includegraphics[width=\textwidth]{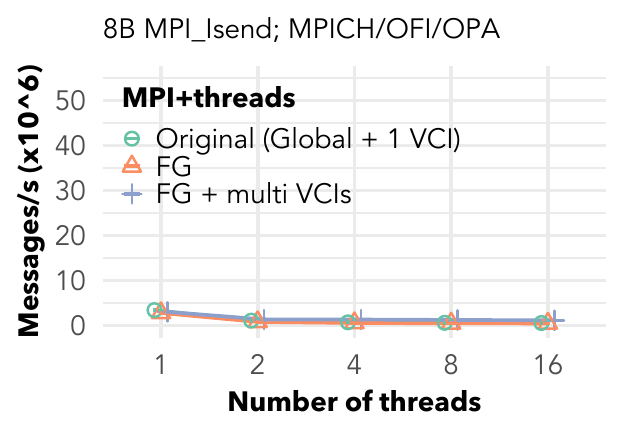}
    \caption{Multiple VCIs.}
    \label{fig:no_opts}
  \end{minipage}
  \begin{minipage}[t]{0.24\textwidth}
    \centering
    \includegraphics[width=\textwidth]{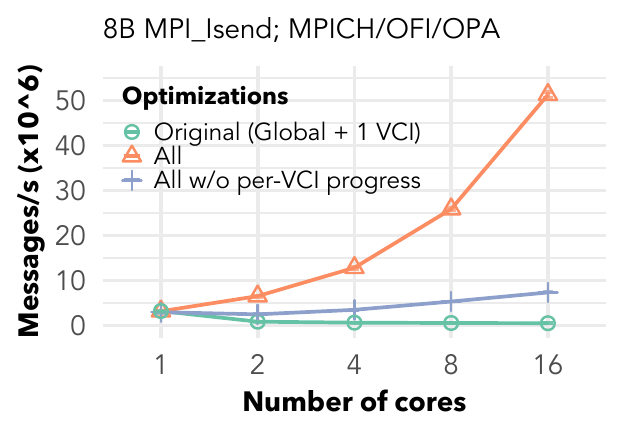}
    \caption{Progress opts.}
    \label{fig:prog_opts}	
  \end{minipage}
  \begin{minipage}[t]{0.24\textwidth}
    \centering
    \includegraphics[width=\textwidth]{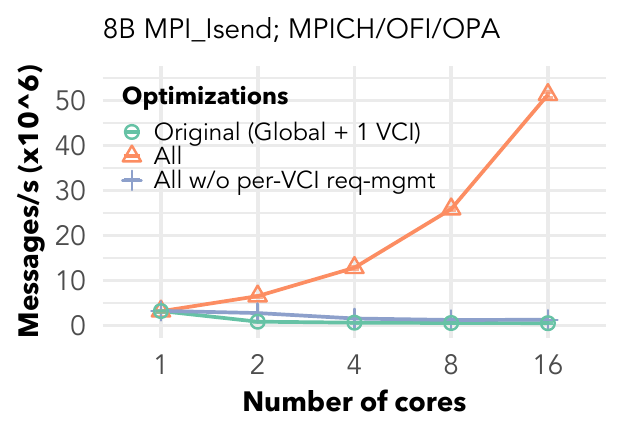}
    \caption{Request opts.}
    \label{fig:req_opts}	
  \end{minipage}
  \begin{minipage}[t]{0.24\textwidth}
    \centering
    \includegraphics[width=\textwidth]{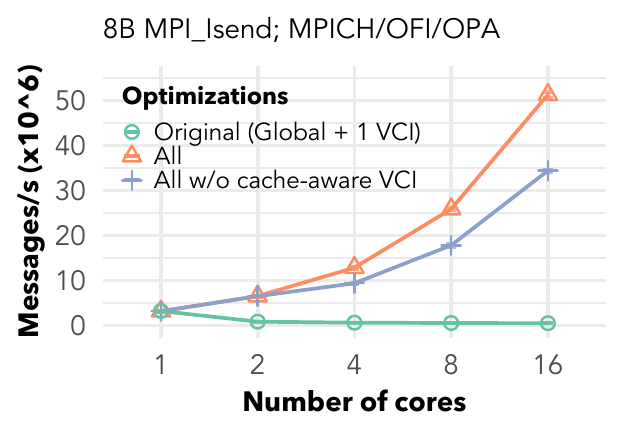}
    \caption{Cache-aware VCI.}
    \label{fig:cache_opts}	
  \end{minipage}
\end{figure*}

\subsection{Optimizing multi-VCI communication}
\label{sec:multi_vci_opts}

\figref{fig:no_opts} shows that simply using multiple VCIs produces
practically no performance benefit. We present several optimizations
to the multi-VCI communication introduced in
Section~\ref{sec:multi_vcis}.

\begin{figure}[t!]
  \begin{minipage}{.238\textwidth}
    \begin{lstlisting}{Pt2pt}
/*Point-to-point example*/
Rank 0:
  MPI_Ssend(comm1);
  MPI_Ssend(comm2);


  
Rank 1 / Thread 0:
  MPI_Irecv(comm1,req1);
#pragma omp barrier
#pragma omp barrier
  MPI_Wait(req1);
  
Rank 1 / Thread 1:
  MPI_Irecv(comm2,req2);
#pragma omp barrier
  MPI_Wait(req2);
#pragma omp barrier
    \end{lstlisting}
  \end{minipage}\hfill
  \begin{minipage}{.238\textwidth}
    \begin{lstlisting}{RMA}
/*RMA example (large Puts)*/
Rank 0:
  MPI_Get(win1);
  MPI_Get(win2);
  MPI_Win_flush(win1);
  MPI_Win_flush(win2);

Rank 1 / Thread 0:
  MPI_Get(win1);
#pragma omp barrier
#pragma omp barrier
  MPI_Win_flush(win1);

Rank 1 / Thread 1:
  MPI_Get(win2);
#pragma omp barrier
  MPI_Win_flush(win2);
#pragma omp barrier
    \end{lstlisting}
  \end{minipage}
  \caption{Point-to-point (left) and RMA (right) scenarios that would
    deadlock without shared progress of VCIs.}
  \label{code:deadlock}
  \vspace{-0.75em}
\end{figure}

\minititle{Per-VCI progress.} With only one VCI (Original), the job of
the progress function was simple: poll for progress on the single
VCI. With multiple VCIs, a \naive\ extension would be to poll for progress on
all the active VCIs. Although correct, this approach would be
detrimental to performance especially when multiple threads progress
operations in parallel since they would be contending on the VCIs'
locks. Also, each thread would be doing more work than
necessary. Because all MPI communication operations map to a VCI,
progress for an operation primarily needs to poll the VCI on which the
operation was posted. We extend the progress engine to allow for
\emph{per-VCI progress}. First, we store the VCI used for an operation
in its request object.  This action adds 3 instructions to the
critical path.  Using the information stored in the request object,
the progress functions poll for progress on the VCI that was used for the
operation. When multiple threads progress operations mapped to
different VCIs, they do not contend.


Although per-VCI progress helps improve performance,
progressing only the VCI used by the current request is incorrect and
can lead to deadlock.  Consider the point-to-point example in
\figref{code:deadlock}.  This is a correct MPI program---the first
synchronous send\footnote{conceptually similar to an
  \mpisend\ following the rendezvous protocol.} on rank 0 (line 3)
should return because its matching receive has already been posted
(line 9).  With current MPI libraries, this program completes because
\mpiwait \texttt{(req2)} (line 17) initiates the reception of
\mpissend \texttt{(comm1)} by polling the single VCI that both
communicators map to, thus allowing \mpissend \texttt{(comm1)} to
return.  With multiple VCIs and per-VCI progress, \mpiwait
\texttt{(req2)} progresses only the VCI associated with
\texttt{comm2}, preventing \mpissend \texttt{(comm1)} to complete and
causing deadlock.  \figref{code:deadlock} also describes a similar
scenario with RMA operations using passive-target synchronization for
cases where the underlying network requires target-side CPU
involvement for progress.

In summary, the pure per-VCI progress model can improve performance,
but the global progress model is necessary to ensure correctness even
though it loses some performance.  To account for such communication
patterns, we use a hybrid progress model; that is, we perform one
round of global progress after a certain number of unsuccessful
per-VCI progress attempts to complete an operation.  We demonstrate
the benefit of our hybrid per-VCI optimization in
\figref{fig:prog_opts}.  Communication throughput is $6.97\times$
lower without per-VCI progress (All w/o per-VCI progress) compared
with the case where all optimizations are used.

\begin{table*}
  \centering
  \caption{Summary of locks on the critical path of initiation and
    progress operations in different critical sections.}
  \label{tab:locks}
  \begin{tabular}{ | r | l | l | l | l | l | }
    \hline
    \textbf{Critical section \textbackslash\ MPI op.} & \textbf{Isend} & \textbf{Isend (immediate)} & \textbf{Put} & \textbf{Wait} & \textbf{Wait (immediate)} \\ \hline		
    \textbf{Global} & 1 (Global) & 1 (Global) & 1 (Global) & 1 (Global) & 1 (Global) \\ \hline
    \textbf{FG} & 2 (VCI + Request) & 1 (VCI) & 1 (VCI) & 2 (VCI + Request) & 0 \\ \hline		
    \textbf{FG + per-VCI req-cache} & 1 (VCI) & 1 (VCI) & 1 (VCI) & 2 (VCI + VCI (request freeing)) & 0 \\ \hline
  \end{tabular}
\end{table*}

\minititle{Per-VCI request management.}  Even when operations from
multiple threads map to different VCIs, they contend on the
request-class's lock when they need to acquire a request (\eg, during
an \mpiisend) or release it (\eg, during an \mpiwait). To address this
contention, we maintain a cache of requests for each VCI. Access to
each cache is protected by the VCI's lock. During the creation of a
request, we first attempt to acquire a request from the cache
belonging to the VCI that the operation maps to. This does not require
acquiring an extra lock because the lock for the VCI is already held
for the operation. If the cache is empty, we fall back on acquiring a
request from the global pool, which requires acquiring the request
class's lock. The caching idea extends to releasing a request to the
cache of a VCI as well. Thus, in the common case, we reduce the number
of lock acquisitions in initiation operations to 1 (FG+per-VCI
req-cache in \tabref{tab:locks}, which summarizes the locks taken in
different critical sections).  Although the request class's lock is
not taken (in the common case) for progress functions either, the
VCI's lock pertaining to the request is taken twice---the final
freeing of the request occurs in the MPI runtime layer, outside the
critical section that protects the progress of the VCI.

In addition to traditional requests, MPICH maintains the pre-completed
lightweight request described in \secref{sec:fg_cs}.  A lightweight
request is a single object and not a pool, so it cannot be cached like
traditional requests. What we do instead is replicate this lightweight
request and provide each VCI with its own. The per-VCI lightweight
requests do not need atomic operations for their updates since each is
protected by the lock of the VCI it belongs to.

\figref{fig:req_opts} shows the benefits of the per-VCI request
management optimizations. Without the optimizations, throughput is
$39.98\times$ lower (All w/o per-VCI req-mgmt) compared with all
optimizations.

\minititle{Cache-line awareness for VCIs}.  We implement the VCI pool
as an array of structs. Each VCI struct holds the lock for that
VCI. Locks of consecutive VCIs are likely to lie on the same cache
line, resulting in the effects of false sharing when threads map to
different VCIs. Hence, we use compiler attributes to cache-align each
VCI. \figref{fig:cache_opts} shows that without a cache-aware VCI, the
message rate is $1.49\times$ lower (All w/o cache-aware VCI).


\minititle{Summary}.  All the thread-safety and multi-VCI
optimizations described in this section are critical for enabling fast
parallel streams of communication for \threads.  The message rate
achieved by the optimized MPI library with 16 threads for 8-byte
\mpiisend{}s is $94.43\times$ higher than that of the state of the
art.

\section{Microbenchmark analysis}
\label{sec:microbench_eval}

We first measure the aggregate message rate of \mpiisend\ and
\mpiput\ (passive target synchronization) using a
communication-intensive benchmark.  The benchmark demonstrates the
maximum rate at which multiple cores can inject messages into the
network simultaneously.  Each core on the host node targets a distinct
core on the remote node. We compare the following modes of execution.

\begin{itemize}

\item \everywhere\ parallelism using the original MPICH version that
  uses one VCI.

\item \threads\ (ser\_comm+orig\_mpich) parallelism with the user not
  exposing communication parallelism on the original MPICH that uses
  one VCI and the Global critical section.

\item \threads\ (ser\_comm+vcis)---same as above but using the
  optimized multi-VCI based MPICH/CH4.

\item \threads\ (par\_comm+orig\_mpich) parallelism with user-exposed
  parallelism on the original MPICH.

\item \threads\ (par\_comm+vcis)---same as above but using the
  optimized multi-VCI based MPICH/CH4.

\item \threads\ parallelism with user-visible endpoints on top of
the optimized multi-VCI infrastructure where each endpoint is a VCI.

\end{itemize}

User-visible endpoints enable explicit control over VCIs, allowing
users to specify the endpoint to use on the host and the remote
endpoint to target. The communication performance of user-visible
endpoints reflects the upper bound of the MPI-3.1 implementation
wherein users implicitly use VCIs through MPI-3.1 mechanisms.

For our analysis with \threads{}, we spawn one rank per node
with an OpenMP thread per core. \everywhere{} uses a rank
per core. In our microbenchmarks with user-visible endpoints, each host-target thread
pair uses its own endpoint, thus exposing communication parallelism to
the MPI library.  With MPI-3.1, when users do not expose parallelism
(ser\_comm), all threads use the same communicator or window.  In
user-exposed parallelism (par\_comm), each thread pair uses its own
communicator or window.

\subsection{Well-behaved communication}
\label{sec:well-behaved}

\begin{figure}[t!]
	\begin{center}
		\includegraphics[width=0.49\textwidth]{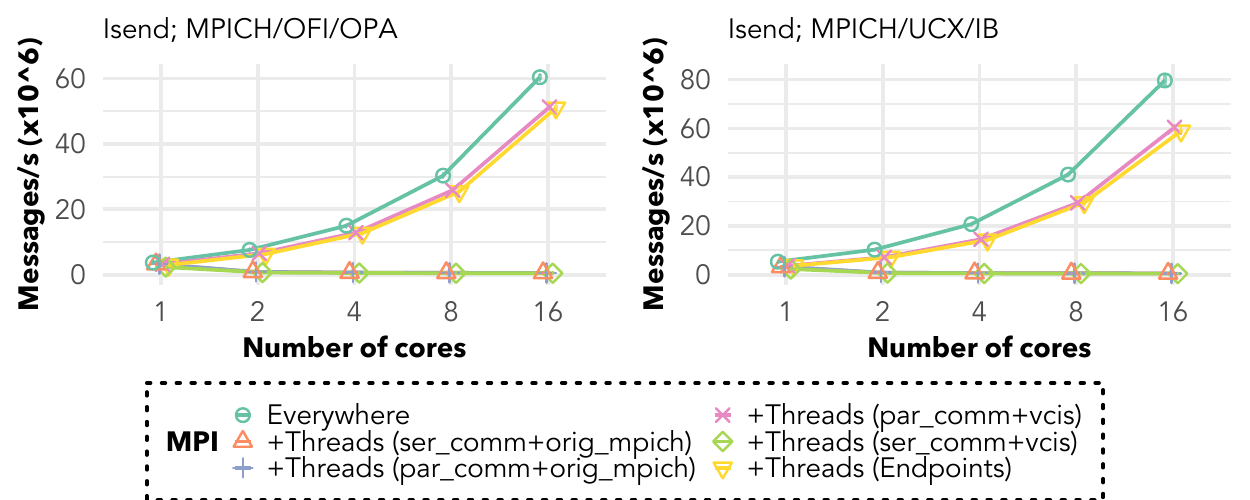}
	\end{center}
	\caption{Message-rate scalability of 8-byte \mpiisend.}
	\label{fig:isend-scaling}
\end{figure}

\begin{figure}[t!]
	\begin{center}
		\includegraphics[width=0.49\textwidth]{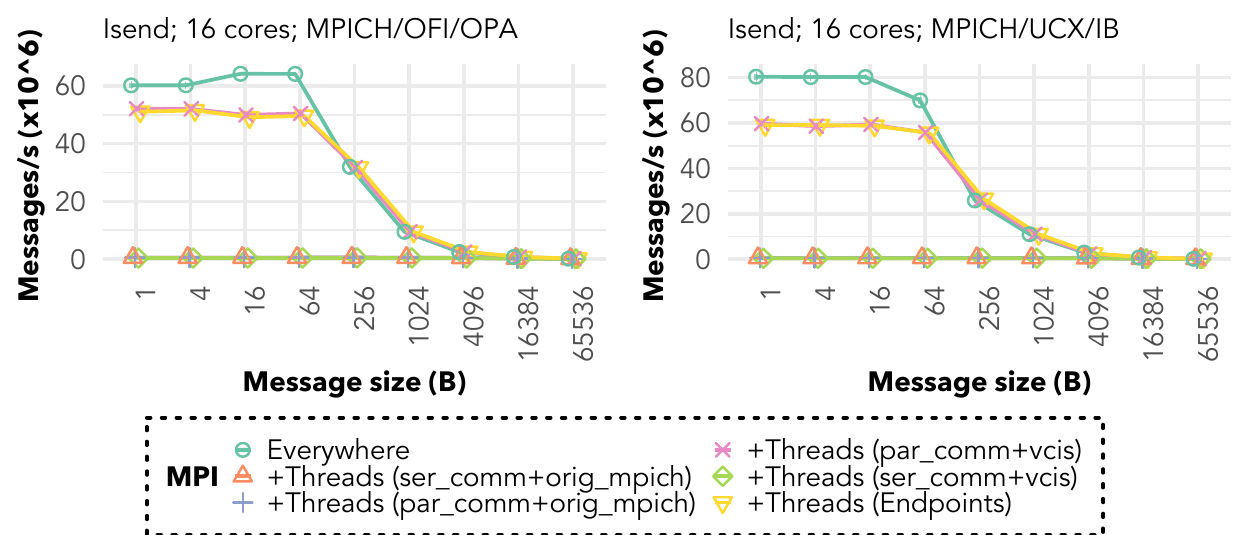}
	\end{center}
	\caption{\mpiisend\ throughput with varying message sizes.}
	\label{fig:isend-varmsgsize}
\end{figure}

For the different modes of execution on OFI/OPA and UCX/IB,
\figref{fig:isend-scaling} shows the message-rate scalability of a
small-message \mpiisend, and \figref{fig:isend-varmsgsize} shows the
message rate of \mpiisend\ with 16 cores across varying message sizes.
\everywhere\ achieves the highest throughput in all cases.  When users
expose communication parallelism, they achieve the same performance
irrespective of the use of VCIs (par\_comm) or user-visible endpoints.
When users expose no communication parallelism (ser\_comm), there is
no performance gain with increasing number of threads.

\begin{figure}[t!]
  \begin{center}
    \includegraphics[width=0.24\textwidth]{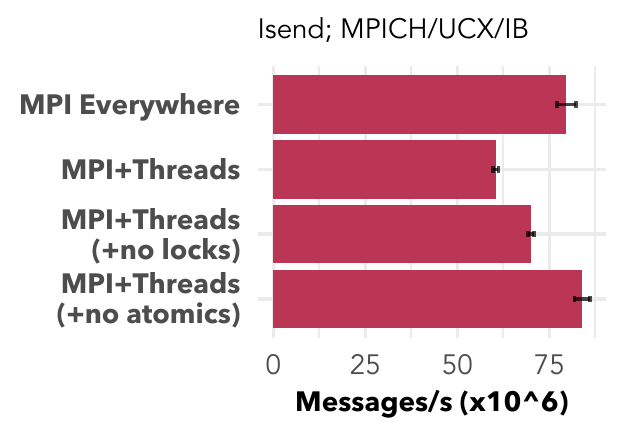}
  \end{center}
  \tiny Note: error bars overlap.
  \vspace{-1em}
  \caption{\threads\ costs.}
  \label{fig:lock_atomic_overheads}
\end{figure}

\minititle{Thread safety costs.} A corresponding
\everywhere\ configuration represents the practical upper bound of the
communication performance of an \threads\ configuration.  Our
optimized \threads\ library utilizes the same level of network
parallelism as \everywhere.  However, \threads\ incurs thread safety
overheads over \everywhere\ even in the uncontended case. These
overheads are most visible for small messages (see
\figref{fig:isend-varmsgsize}) since the message rate is bound by the
CPU, not by the network.  The sources of the thread safety overheads
are lock acquisitions and atomics for completion or reference
counting.
\figref{fig:lock_atomic_overheads} shows that
if we disable locking and atomics,\footnote{Since each thread maps to
	its own VCI in the \threads\ microbenchmark, disabling thread
	safety, although incorrect, does not lead to erroneous behavior.}
\threads\ can match the throughput of \everywhere. One solution to
mitigate thread-safety costs would be to allow users to provide hints
as to which communicators will be accessed by a dedicated thread,
thereby allowing the MPI library to disable locking for the VCIs of 
communicators accessed by dedicated threads.

\begin{take}
	\emph{Takeaway:} For basic communication, VCIs and endpoints perform
	similarly and nearly as well as \everywhere.
\end{take}

\subsection{Not-so-well-behaved communication}
\label{sec:rma-comm}

\begin{figure}[t!]
	\begin{center}
		\includegraphics[width=0.49\textwidth]{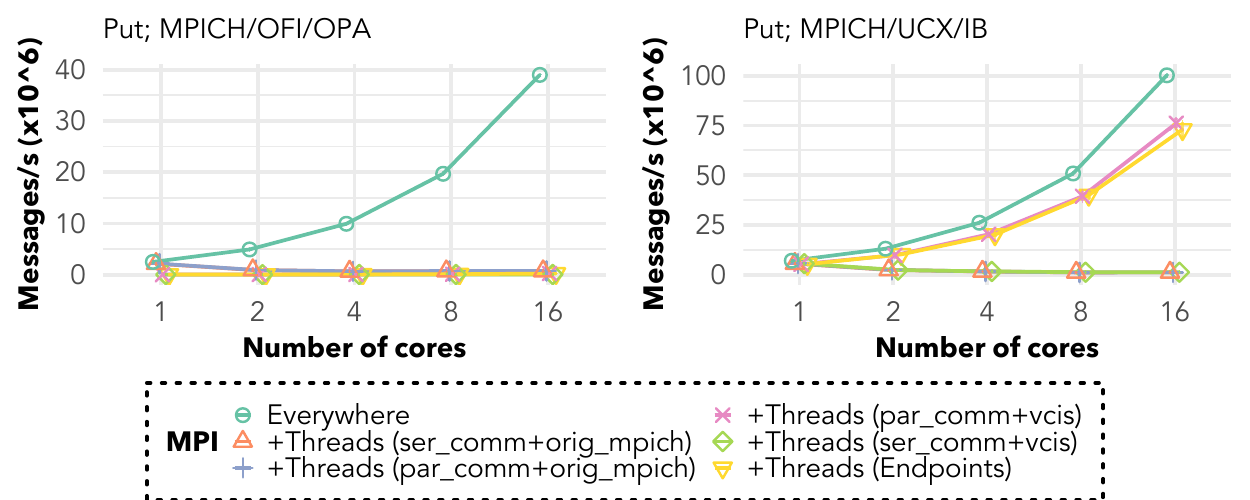}
	\end{center}
	\caption{Message-rate scalability of 8-byte \mpiput.}
	\label{fig:put-scaling}
\end{figure}

Similar to \figsref{fig:isend-scaling}{fig:isend-varmsgsize},
\figsref{fig:put-scaling}{fig:put-varmsgsize} demonstrate, for the
different modes of execution on OFI/OPA and UCX/IB, the throughput
scalability of a small-message \mpiput, and the 16-core message rate
of \mpiput\ across varying message sizes, respectively

\minititle{Network hardware limitations.} The \threads\ message rate
of \mpiput\ on OFI/OPA is dismal even with exposed parallelism on VCIs
and user-visible endpoints.  The reason is that Intel OPA emulates its
RMA operations in software, requiring the application on the target
side to actively progress a VCI for a performance-oriented execution
of the operation.  When the application provides no help, OPA relies
on its low-frequency PSM2 progress thread for completion of the
operation.  In our benchmark, all the threads from all processes first
initiate their RMA operations in parallel.  Then, one thread waits on
an MPI barrier, after which all threads synchronize with a thread
barrier.  The communicator used for the MPI barrier internally uses a
VCI different from those of the windows on which the RMA operations
are issued.  Thus, none of the threads directly make progress on the
incoming messages of the RMA VCIs.  The thread waiting on the MPI
barrier occasionally performs global progress, so the benchmark
eventually completes, but such global progress is infrequent and thus
hurts performance.

With UCX/IB, on the other hand, we see no such degradation in
performance because Mellanox IB is capable of implementing contiguous
\mpiput\ operations fully in hardware.  Thus, even if the target
threads are not making direct progress on the RMA VCIs, the operations
still complete quickly.

\begin{figure}[t!]
	\begin{center}
		\includegraphics[width=0.49\textwidth]{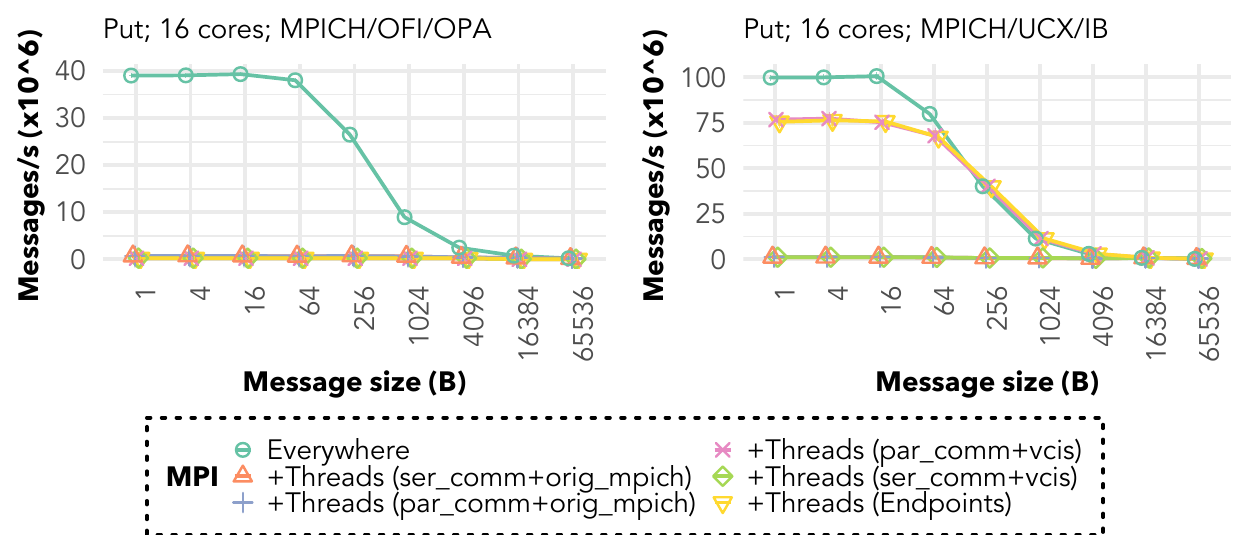}
	\end{center}
	\caption{\mpiput\ throughput with varying message sizes.}
	\label{fig:put-varmsgsize}
\end{figure}

\begin{figure}[t!]
	\begin{center}
		\includegraphics[width=0.49\textwidth]{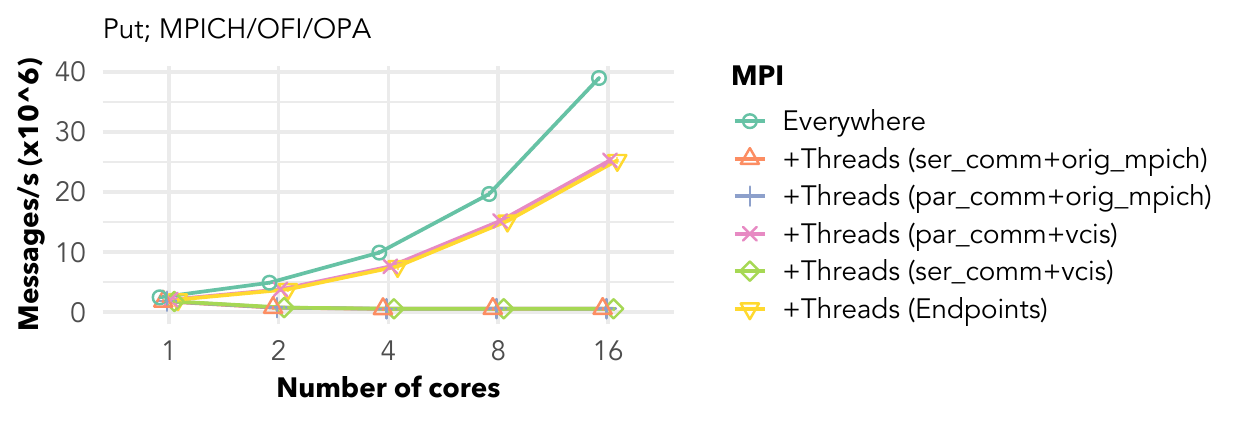}
	\end{center}
	\caption{Parallel Win\_free.}
	\label{fig:par_win_free}
\end{figure}

With \everywhere{}, each process has a single VCI.  Thus, the target
ranks waiting on an MPI barrier continuously progress the VCI being
targeted by the initiator ranks.

The main point demonstrated here is the tradeoff between dedicated
progress and shared progress. \everywhere{} has no distinction between
dedicated and shared progress because it only has a single VCI.  For
\threads{}, when a single VCI is used (\ie, original MPICH), like
\everywhere{}, it has no distinction between dedicated and shared
progress either.  But, for \threads{}, when we use multiple VCIs, the
same independence of VCIs that enables good performance through the
avoidance of locks also hurts shared progress between the threads.
One can work around this issue by, for example, having each thread be
responsible for progress on its window (in the same way that
\everywhere\ works).  One possibility is that threads call
\mpiwinfree\ on their own windows in parallel (see
\figref{fig:par_win_free}), thus making progress on the corresponding
VCIs, although how practical this possibility is in real applications
remains to be seen.

\begin{figure}[t!]
	\centering
	\begin{minipage}[t]{0.236\textwidth}
		\centering
		\includegraphics[width=\textwidth]{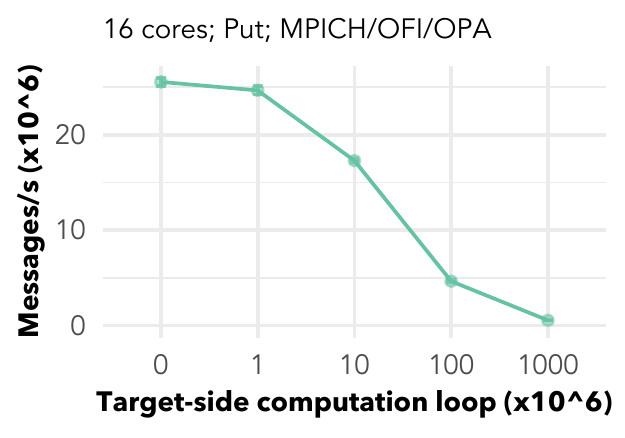}
		\caption{Busy target.}
		\label{fig:multiple-var-target-comp}
	\end{minipage}
	\begin{minipage}[t]{0.236\textwidth}
		\centering
		\includegraphics[width=\textwidth]{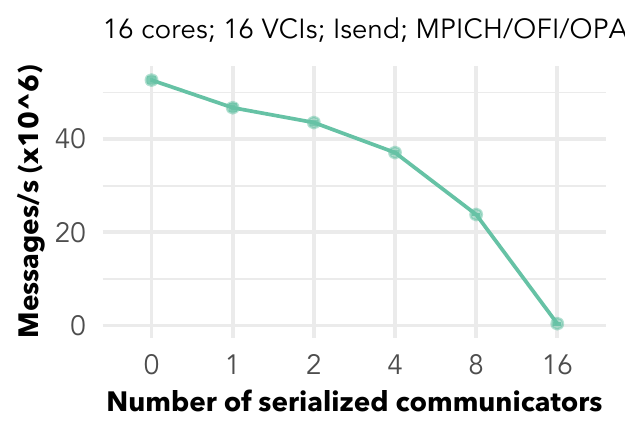}
		\caption{Mapping mismatch}
		\label{fig:var-serial-comms}
	\end{minipage}
     \vspace{-1.5em}
\end{figure}

\minititle{Busy target.} Typically, the target side is involved in its
own computational activities and does not just wait for communication
to complete, as in \figref{fig:par_win_free}. The target's computation
then determines the productivity of operations that need the target
VCI to be progressed. \figref{fig:multiple-var-target-comp} shows a
deteriorating \mpiput\ message rate when the computation before the
call to \mpiwinfree\ increases on the threads of the target rank.

\begin{take}
	\emph{Takeaway:} When shared progress is required neither VCIs nor
	user-visible endpoints perform well.
\end{take}

\minititle{Mismatch in expected mapping to VCI.}  Even if the user
exposes parallelism, parallel operations can contend on the same VCI
because the number of VCIs available is hardware dependent and
typically small, and these VCIs themselves are not exposed to the user.
A simple first-come, first-served model of VCI allocation to
communicators might not be the best strategy in this regard because
the user cannot identify which communicators map to distinct VCIs and
can therefore be used by different threads for better performance.
\figref{fig:var-serial-comms} shows
the deteriorating effect on throughput for increasing amounts of
serialization when there are 16 threads and the network hardware
features only 16 contexts.  The user is exposing communication
parallelism, but the observed performance is low because of the
mismatch in expectations of mapping to the underlying VCIs.
One solution to this mismatch in expected mapping would be to
allow users to provide hints as to which communicators are used
by different threads and can benefit from independent VCIs.

User-visible endpoints, compared to MPI-3.1, can perform better in this
situation because they expose the network hardware contexts to the
user.  We note that user-visible endpoints could also be implemented
as a virtual layer on top of internal VCIs. However, they form a closer
mapping to network resources than what communicators do.  

\begin{take}
	\emph{Takeaway:} User-visible endpoints allow users to carefully
	manage their communication, thus performing better in
	situations when the MPI-3.1 library serializes user-exposed parallelism.
\end{take}

\subsection{Limiting MPI semantics}
\label{sec:mpi-limit}

\begin{figure}[t!]
	\centering
	\begin{minipage}[t]{0.236\textwidth}
		\centering
		\includegraphics[width=\textwidth]{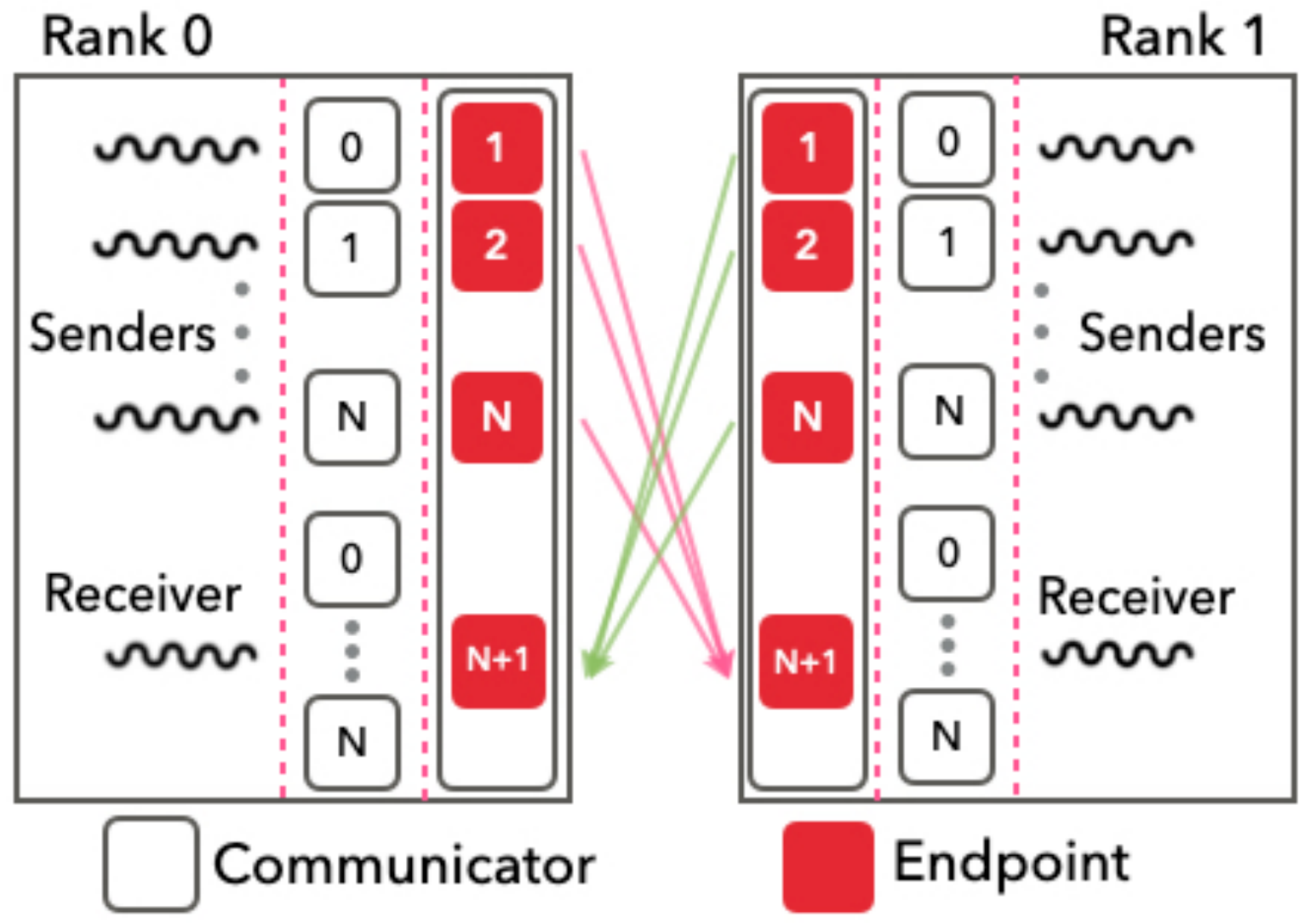}
		\caption{Dedicated threads.}
		\label{fig:dedicated-thread-communication}
	\end{minipage}
	\begin{minipage}[t]{0.236\textwidth}
		\centering
		\includegraphics[width=\textwidth]{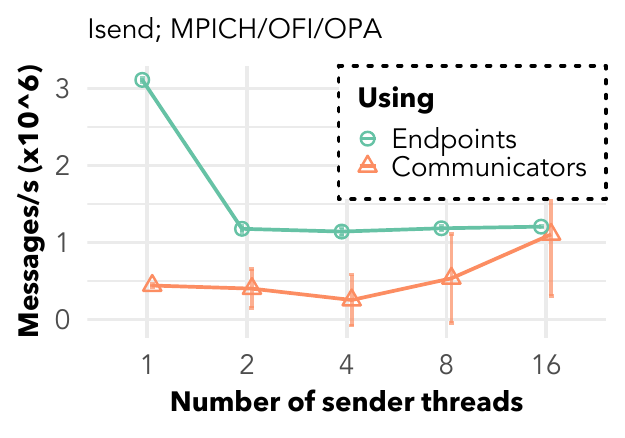}
		\caption{Hurtful abstraction}
		\label{fig:multi-sender-single-receiver}
	\end{minipage}
	\vspace{-1.5em}
\end{figure}

Abstracting the use of VCIs through communicators can hurt certain
irregular communication
patterns. \figref{fig:dedicated-thread-communication} captures the
communication pattern of Legion's~\cite{bauer2012legion} runtime,
which maintains a set of threads where a few are dominant message
senders and a few are dedicated polling threads that receive
messages. With MPI-3.1, each sender thread uses a separate
communicator. However, the semantics of MPI require the receiver
thread to iterate over the communicators, thus forcing the receiver to
contend on the VCIs of the senders and hurting performance (see
\figref{fig:multi-sender-single-receiver}).  With user-visible
endpoints, this contention does not exist since each thread uses a
distinct endpoint and can directly address the endpoint of the remote
receiver. The single receiver is a bottleneck with both user-visible
endpoints and communicators. With communicators, the fraction of time
spent by the receiver on a VCI's lock decreases with increasing number
of senders.  Hence, its performance approaches that with endpoints.

\begin{take}
  \emph{Takeaway:} When MPI's semantics limit the user from exposing
  parallelism, user-visible endpoints perform better than VCIs.
\end{take}

\section{Application Analysis}
\label{sec:miniapps}

In this section, we showcase the communication performance of three
applications, one from each of the three categories described in
\secref{sec:introduction}. For each application, we compare the
performances of \everywhere\ (a rank per core) and \threads\ (a
rank per node; an OpenMP thread per core) parallelism. For \threads ,
we show the performances of user-exposed parallelism on VCIs and on
the original MPI library and compare them with that of user-visible
endpoints.

\subsection{Stencil applications}
\label{sec:stencil}

Stencils are arguably the most common design patterns in HPC
applications.  They are at the heart of various application domains
such as computational fluid dynamics, image processing, and partial
differential equation solvers.  Prominent applications with the
stencil communication pattern include Nek5000~\cite{nek5000-web-page}
and LAMMPS~\cite{plimpton1995fast}.

\begin{figure}[t!]
	\centering
	\begin{minipage}[t]{0.236\textwidth}
		\centering
		\includegraphics[width=\textwidth]{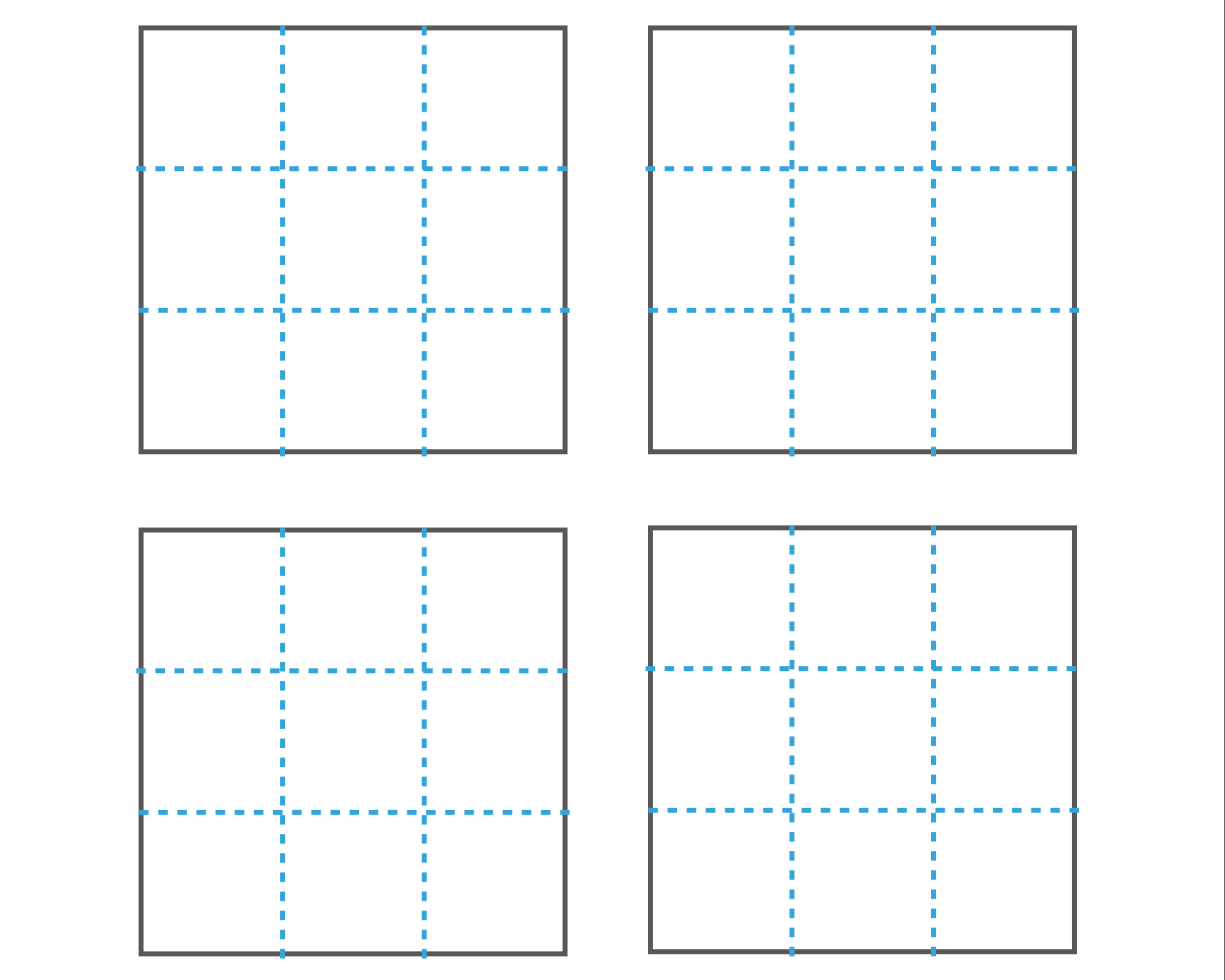}
		\captionof{figure}{6x6 grid with 3x3 sub-blocks per node.}
		\label{fig:stencil-partitioning}
	\end{minipage}
	\begin{minipage}[t]{0.236\textwidth}
		\centering
		\includegraphics[width=\textwidth]{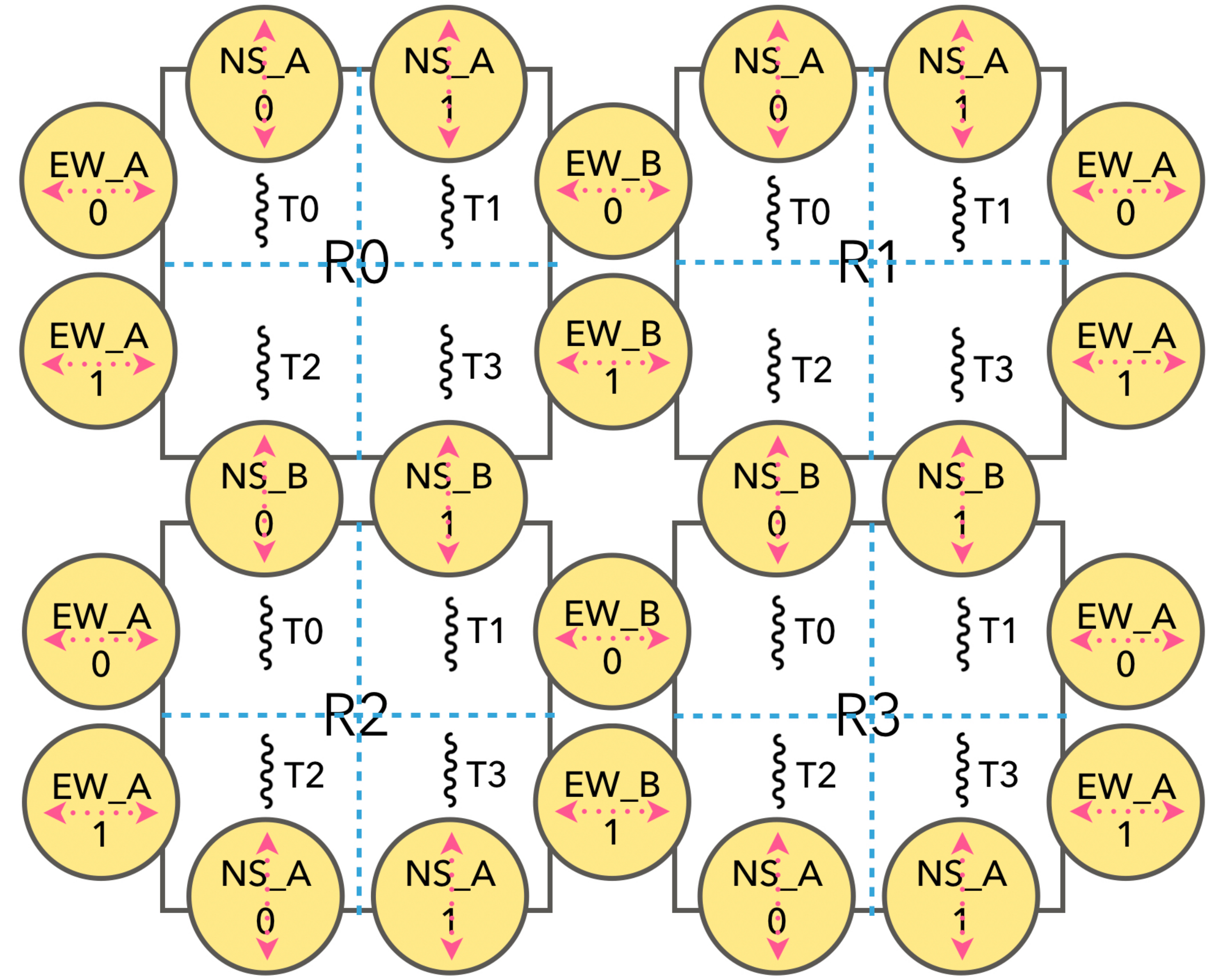}
		\captionof{figure}{Logical parallelism in \threads{} stencil.}
		\label{fig:stencil-comm-usage}
	\end{minipage}
	\vspace{-1.5em}
\end{figure}

Using a 2D 5-point stencil, we evaluate the neighborhood halo exchange
(non-blocking point-to-point) time per iteration of the stencil
pattern. We first partition the mesh into blocks across nodes, and
then within each node we further partition the sub-block among cores
(\figref{fig:stencil-partitioning} shows an example). The squares
formed by the intersection of the dashed blue lines represent cores
that are driven by processes and threads in \everywhere\ and
\threads\ parallelism, respectively. The blue dashed lines also
represent boundaries where the halo exchange takes place through
shared memory.  MPI still executes intranode halo exchanges in
\everywhere.  In \threads, threads use MPI only for internode halo
exchanges and directly read the shared memory for intranode
communication.  The stencil pattern falls into the first category of
applications---the internode communication of threads on edges of the
nodes is independent and can execute on its own communication stream.

With user-visible endpoints, we create as many endpoints as there are
threads on edges. For the example in
\figref{fig:stencil-partitioning}, we create 8 endpoints per
node. Each communicating thread uses its own endpoint and exchanges
halos by addressing the ranks of the remote endpoints, thereby
achieving parallel communication.  With MPI-3.1, we use two sets of
communicators---odd and even---for each of the north-south and
east-west exchanges. Each set contains as many communicators as there
are threads on the node edge.  \figref{fig:stencil-comm-usage} shows
an example.  Depending on the Cartesian coordinates of the rank, the
threads on a rank would use either the odd set or the even set.  The
odd-even sets prevent multiple threads from using the same
communicator. Without them, T0 on R0 and R2 in
\figref{fig:stencil-comm-usage} would use the same NS\_0 communicator,
requiring T2 on R0 to also use the NS\_0 communicator and thus
serializing the communication of T0 and T2 on R0. Periodic stencils
where the number of ranks along a dimension of the process-grid is odd
require a separate set of communicators for the wraparound. The
communicator usage can indeed be reduced without hurting performance
by using only one communicator for the threads on corners, since their
halo exchanges execute in serial.

The above example also demonstrates that the matching semantics of
communicators or tags (i.e., the same communicator and tag must be
used for both the sender and receiver) sometimes makes exposing
communication parallelism with MPI-3.1 clumsy compared with that of
user-visible endpoints.  While not a performance argument,
one might consider it to be a productivity concern.

\begin{figure}[t!]
	\begin{center}
		\includegraphics[width=0.49\textwidth]{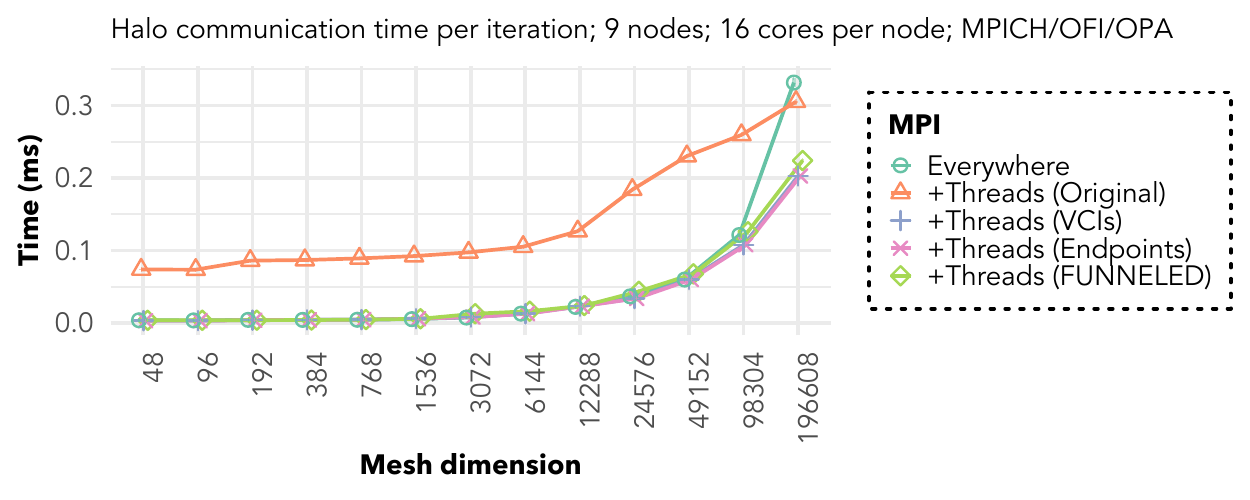}
	\end{center}
	\caption{Halo communication across varying mesh sizes.}
	\label{fig:stencil-eval}
	\vspace{-1.5em}
\end{figure}

Our evaluation utilizes all 9 nodes of the OFI/OPA cluster and engages
16 cores per node. \figref{fig:stencil-eval} shows the halo
communication time\footnote{For the \funneled\ mode, we do not report
  the time spent in packing and unpacking the buffer.} for each mode
across varying mesh dimensions. This time discards the cost of any
load imbalance since we use MPI barriers before the start of each halo
exchange.  We observe that the communication performance of VCIs with
user-exposed parallelism matches that of \everywhere\ parallelism,
user-visible endpoints, and \funneled{}.

\begin{recommendation}
  \emph{Recommendation:} Maximize independence between threads for
  point-to-point communication with MPI communicators.
\end{recommendation}

\begin{warning}
  \emph{Warning:} Independent communication with MPI ranks or tags is
  not sufficient because of wildcards on the receive side.
\end{warning}

\begin{warning}
  \emph{Warning:} The matching requirements of communicators or tags
  sometimes makes exposing communication parallelism with MPI-3.1
  clumsy compared with that of user-visible endpoints.
\end{warning}

\subsection{OpenMC}
\label{sec:ebms}

The Center for Exascale Simulation of Advanced Reactors (CESAR) was a
DOE co-design center whose primary objective was to adapt algorithms
to the next-generation HPC architectures on the path to exascale
systems.  CESAR focused on algorithms that target the high-fidelity
analysis of nuclear reactors. These include algorithms governing
thermal hydraulics and neutronics. Applications simulating the former
typically have a neighborhood, stencil style of communication, which
we evaluated in \secref{sec:stencil}. The latter consists of
distributed Monte Carlo (MC) neutron-transport codes, such as
OpenMC~\cite{romano2014openmc}. Siegel et
al.~\cite{siegel2014improved} presented the original energy-banding
(EB) algorithm for OpenMC, and Felker et al.~\cite{felker2014energy}
extended the EB idea to distributed-memory machines by distributing
the cross-section data (composed of energy bands) across multiple
nodes.  Rather than the domain, particles are evenly distributed
between the nodes.  During simulation, each node fetches one band of
the cross section using \mpiget\ operations, tracks the movement of
its share of particles, and iterates over the number of bands.

CESAR's EBMS miniapp~\cite{ebmsmini} captures the communication
pattern of the distributed EB idea.  It utilizes
\shared~\cite{hoefler2013mpi+}: multiple processes on a node that
share a receive buffer that is large enough to hold one band of the
cross-section.  While the computation is distributed among the
different processes on the node, only one process is responsible for
communication.  We extended the EBMS miniapp to distribute the
communication workload among the processes as well~\cite{rzambre_ebms}. We also
implemented a \threads\ version of the miniapp with one multithreaded
process per node.  The communication workload between the cores is the
same for both the \everywhere\ (+ shared memory) and the
\threads\ versions.

\begin{figure}[t!]
	\begin{center}
		\includegraphics[width=0.5\textwidth]{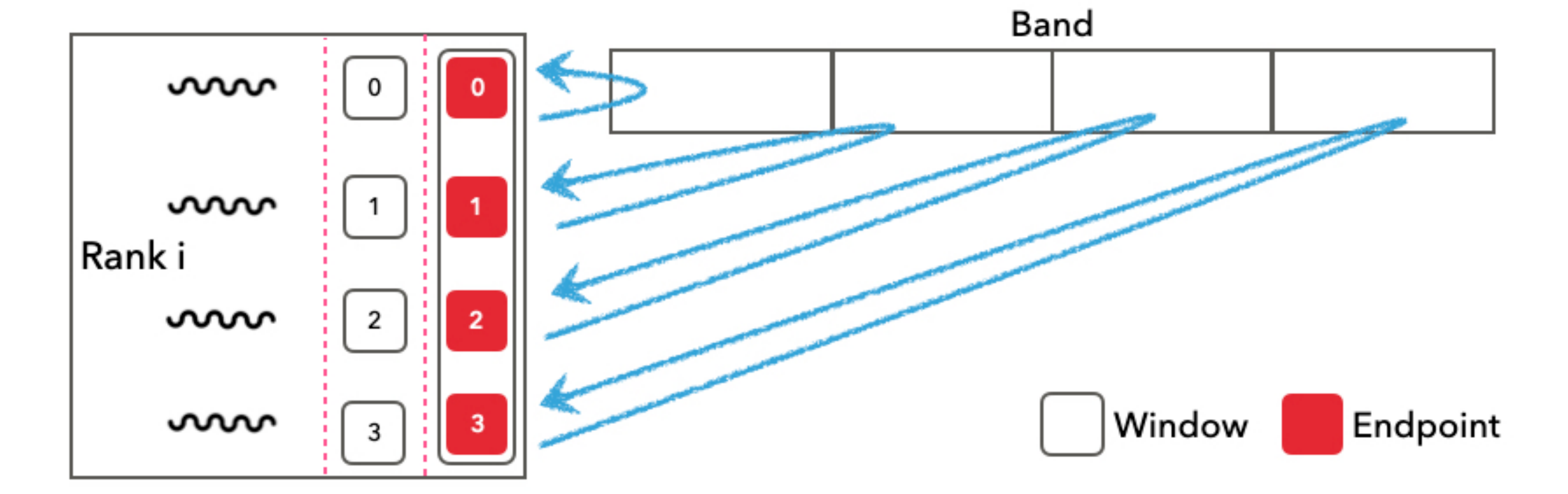}
	\end{center}
	\caption{Logical parallelism in \threads\ EBMS.}
	\label{fig:ebms_pattern}
\end{figure}

\begin{figure}[t!]
	\begin{center}
		\includegraphics[width=0.49\textwidth]{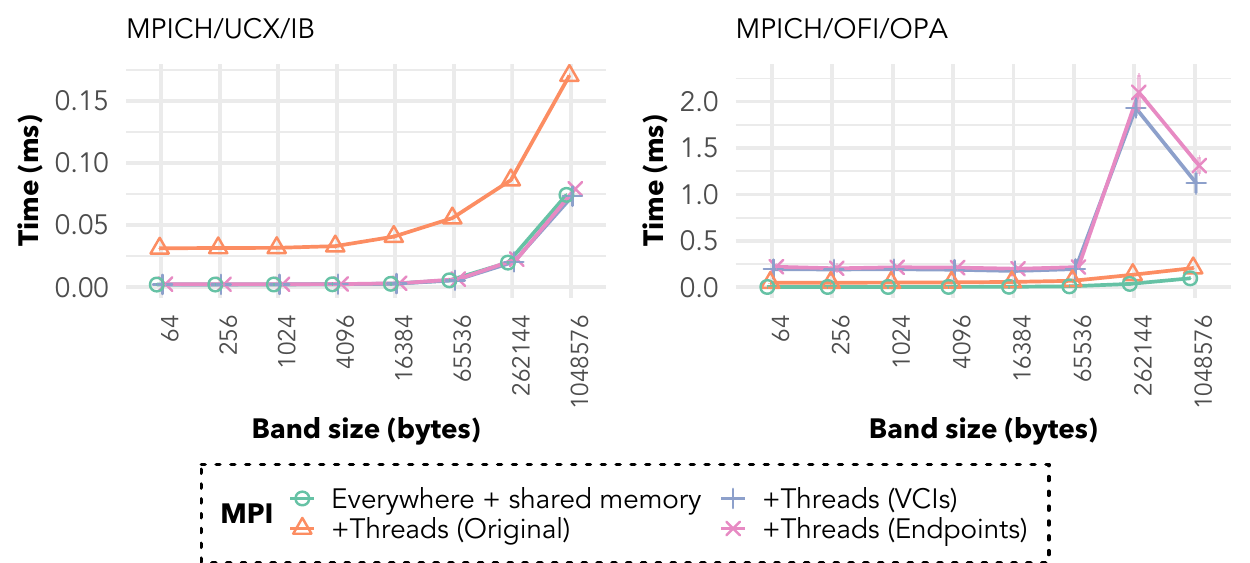}
	\end{center}
	\caption{Time per remote fetch across varying band sizes with 16
		cores per node on UCX/IB (left) and OFI/OPA (right).}
	\label{fig:skylake_gomez_ebms}
\end{figure}

\begin{figure}[t!]
	\begin{center}
		\includegraphics[width=0.49\textwidth]{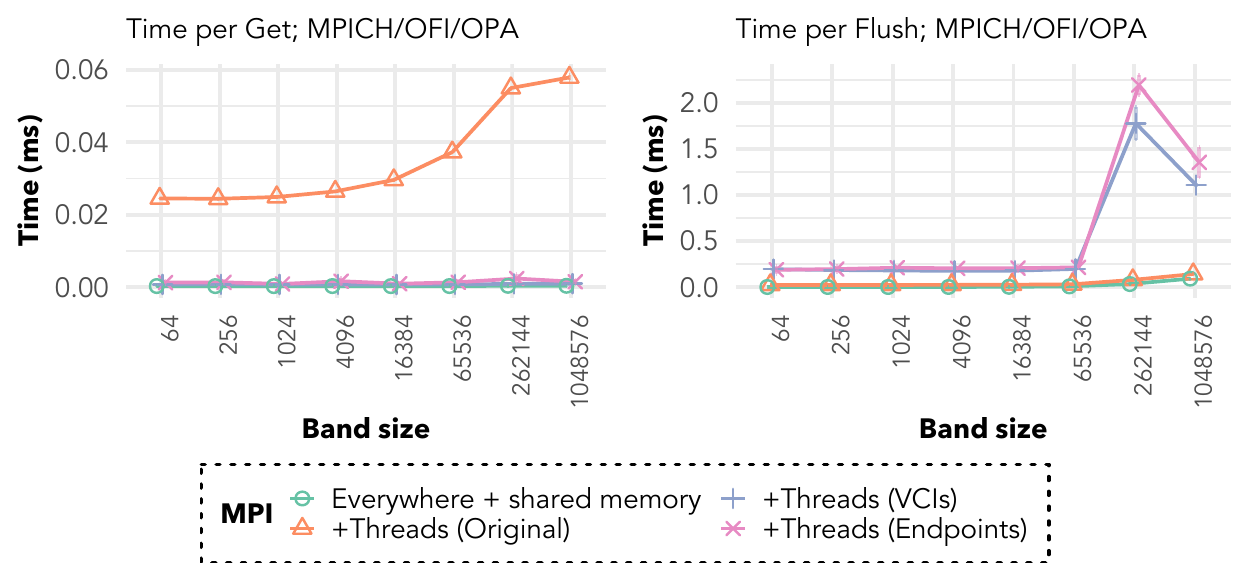}
	\end{center}
	\caption{Get and flush time across varying band sizes on OFI/OPA.}
	\label{fig:get-init-flush-time}
\end{figure}

The EBMS pattern falls into both the first and second categories of
applications (listed in \secref{sec:introduction}).  It falls into the
first category because \mpiget\ operations of different threads are
independent; they can execute on distinct communication streams.  The
pattern falls into the second category because of the use of RMA---the
underlying interconnect may be limited and rely on shared progress.

To leverage the independence between threads with user-visible
endpoints, we create a separate endpoint for each thread.  With
MPI-3.1, we use a separate window per thread as shown in \figref{fig:ebms_pattern}.
The memory is not duplicated for each window.

Our evaluation utilizes 4 nodes and engages 16 cores per node on both
the UCX/IB and OFI/OPA clusters.  We measure the time for each fetch
of a portion of a band that resides on a remote node.  A remote fetch
includes an \mpiget\ and an \mpiflush.
\figref{fig:skylake_gomez_ebms} shows the time for a remote fetch on
the UCX/IB cluster.  The communication performance of \threads\ with
VCIs is the same as that of \everywhere\ and user-visible endpoints.

\begin{recommendation}
  \emph{Recommendation:} Maximize independence between threads for RMA
  communication with MPI windows.
\end{recommendation}

\begin{sloppypar}
On the other hand, the remote-fetch times on OFI/OPA (see
\figref{fig:skylake_gomez_ebms}) show that exposing parallelism on
VCIs hurts performance, especially for large messages.  The case is
the same with user-visible endpoints.  The time for a remote fetch is
governed by the issue of the fetch (\mpiget) and its completion
(\mpiflush).  If we separate them out,
\figref{fig:get-init-flush-time} shows that the time for an
\mpiget\ using multiple VCIs is the same as that in \everywhere\ but
the time of \mpiflush\ is more expensive.  The reason is that the
communication pattern of the application does not guarantee that the
remote VCI being targeted by the \mpiget\ operations will be progressed---the
thread mapped to the target VCI on the remote rank could be waiting on
a thread-barrier that exists between each iteration of the simulation.
Intel OPA relies on the application to make progress on the target VCI
for the completion of large-message RMA transfers and for a productive
execution of small to medium message transfers.  Hence, the execution
is dependent on the occasional global progress in the progress engine.
\end{sloppypar}

\begin{warning}
  \emph{Warning:} Independent communication with VCIs fundamentally
  opposes shared progress.
\end{warning}

\subsection{NWChem}
\label{sec:bspmm}

\begin{figure}[t!]
	\begin{center}
		\includegraphics[width=0.5\textwidth]{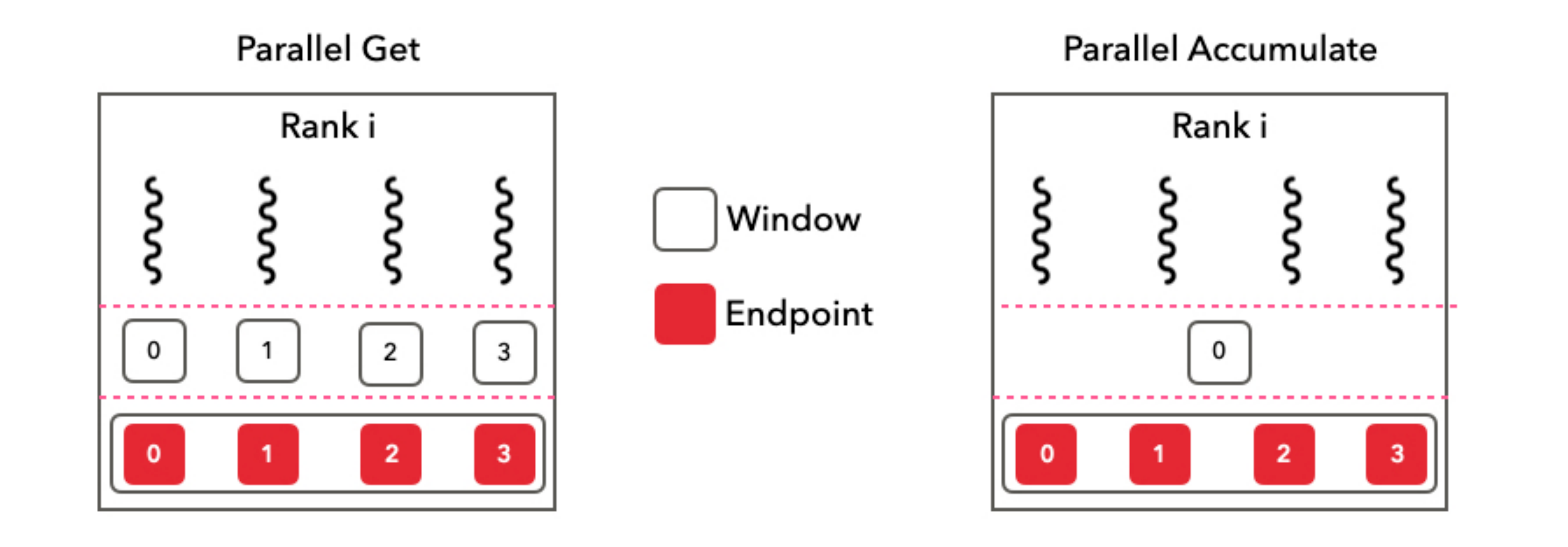}
	\end{center}
	\caption{Logical parallelism in \threads\ BSPMM.}
	\vspace{-0.5em}
	\label{fig:bspmm_pattern}
\end{figure}

NWChem~\cite{valiev2010nwchem} is a prominent quantum chemistry
application suite for large-scale simulations of chemical and
biological systems.  It uses the Global Arrays
(GA)~\cite{nieplocha1994global} library to distribute the
multidimensional arrays across the memories of multiple nodes and
provide access to the data through one-sided MPI operations.  When
NWChem is used for quantum chemical many-body methods, such as CCSD
and CCSD(T), the dominant cost is that of BSPMM: block-sparse matrix
multiplication (tensor contractions). NWChem implements this with
dense matrix operations using a \emph{get-compute-update} pattern:
each worker (processing entity) uses \mpiget\ to retrieve the
submatrices it needs, and after the multiplication it uses an
\mpiaccum\ to update the memory at the target location.

Using a mini-app~\cite{rzambre_bspmm},
we evaluate a 2D version of this communication pattern that performs $A
\times B = C $, wherein the input matrices $A$ and $B$ are composed of
tiles. Each tile is either a dense or zero matrix. The nonzero tiles
are evenly distributed among the ranks in a round-robin fashion. Each
rank maintains a work-unit table that lists all the multiplication
operations that workers need in order to cooperatively execute.  Rank
0 hosts a global counter, which the workers fetch and add atomically
(\mpifetchandop). The fetched counter serves as an index to the
work-unit table.  Each worker locally accumulates its $C$ tiles until
the next fetched work unit corresponds to a different $C$ tile, in
which case the worker uses an \mpiaccum\ to update the $C$ tile. A
worker is a process in \everywhere\ and a thread in \threads.

\begin{figure*}[t!]
	\begin{center}
		\includegraphics[width=0.99\textwidth]{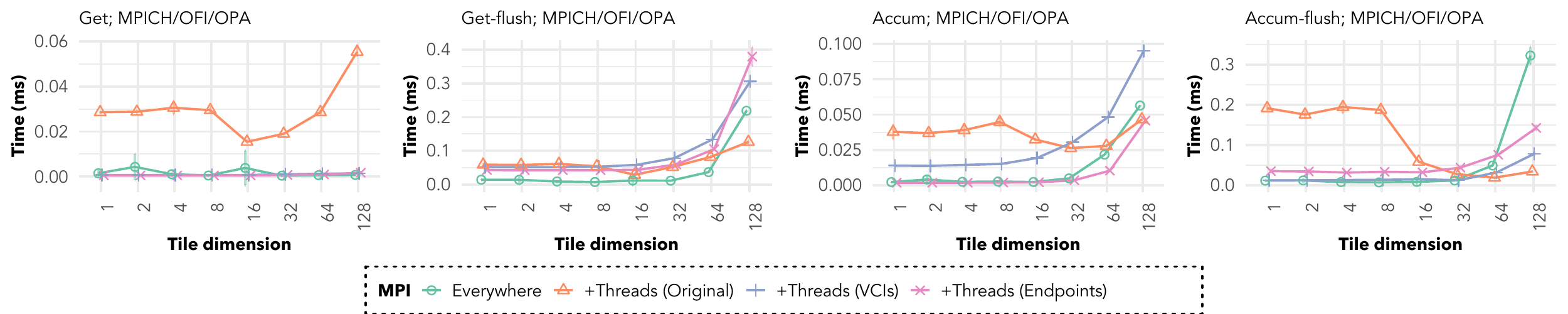}
	\end{center}
	\caption{BSPMM communication performance on Intel Omni-Path.}
	\label{fig:skylake_bspmm}
	\vspace{-0.5em}
\end{figure*}

\begin{sloppypar}
\threads\ BSPMM falls under the third category of applications.
Although each thread can use its own window for its \mpiget\ to fetch
tiles of $A$ and $B$, MPI-3.1's semantics constrain the threads within
a rank to use a single window for the \mpiaccum .  Each thread cannot
use its own window for \mpiaccum\ because atomicity across windows for
the same memory location is undefined. On the other hand, user-visible
endpoints enable the creation of multiple endpoints within a single
window. Hence, each thread uses its own endpoint for both \mpiget\ and
\mpiaccum \ as shown in \figref{fig:bspmm_pattern}.
\end{sloppypar}

\figref{fig:skylake_bspmm} portrays the performance of BSPMM's
communication pattern on 4 nodes of the OFI/OPA cluster with 16 cores
engaged per node.  We measure the time taken to initiate the
operations (\eg, \mpiget) separately from the time taken to complete
them (\eg, \mpiflush). VCIs initiate \mpiget{} operations as fast as
endpoints and \everywhere.  However, only endpoints initiate
\mpiaccum\ operations as fast as \everywhere; \threads\ with MPI-3.1
is constrained by the use of a single window. The flush of
\mpiget\ operations demonstrates behavior similar to that in the EBMS
pattern (see \secref{sec:ebms}).  \threads\ with VCIs flushes
\mpiaccum\ operations faster than endpoints because of its use of a
single VCI---the probability of the remote target VCI being progressed
is higher since all threads on the target rank map to it. The
Accum-flush of \everywhere\ is the slowest for large tile dimensions
because the worker cannot progress its VCI until it finishes its
computational tasks, which is larger for large tile dimensions. On the
other hand, if a worker in \threads\ is busy with computational tasks,
other workers on the same rank might progress the VCI, either because
they all map to the same VCI or because of shared progress, allowing
for a more productive execution of a large-message RMA operation than
that in \everywhere.

\begin{warning}
  \emph{Warning:} Atomic operation semantics are not easy to achieve
  with multiple windows; using multiple VCIs may not help.
\end{warning}

An important point to note is that the \mpiaccum{} operations in BSPMM
do not need to be ordered. Hence, if the user hints this relaxation using
the \texttt{accumulate\_ordering=none} hint, the MPI library
could issue the operations from different threads in parallel and
thereby achieve the same performance as  user-visible endpoints.
Furthermore, increasing the number of ranks per
node and decreasing the number of threads per rank in hybrid \threads{}
could also help the case of accumulates with VCIs. The optimal combination
of ranks per node and threads per rank, however, depends on an empirical
study with the application.
\section{Relevance to MPI-4.0}
\label{sec:discussion}

The next iteration of the MPI standard, MPI-4.0~\cite{mpi4draft},
is considering featuring new info hints (\eg{ \texttt{mpi\_assert\_no\_any\_tag}})
that would provide the users with more opportunities
to expose communication parallelism in their \threads{}
communication. For example, if an application hints that it
does not use wildcards in its communication, MPI-4.0 would
allow the user to expose communication parallelism through
tags within a single communicator in addition to the option of exposing
parallelism through communicators. These new ways to expose
parallelism would, in turn, need to be mapped to the multiple VCIs
inside the MPI library. Hence, the productive use of the new
hints relies on the muti-VCI infrastructure that this work provides.
\section{Related Work}
\label{sec:related_work}

The communication performance of \threads\ has been a decade-long
concern. Researchers have studied the problem in various ways, ranging
from mitigating lock contention on the MPI library's software
resources~\cite{amer2015mpi+,amer2019software,balaji2010fine} to
extending the MPI
standard~\cite{dinan2013enabling,grant2019finepoints}.  We discuss
prior works that are conceptually related to ours.

\subsection{MPI Endpoints demonstration}

Dinan et al.~\cite{dinan2014enabling} and Sridharan et
al.~\cite{sridharan2014enabling} demonstrate the performance of
\threads\ with the MPI Endpoints proposal. Although they agree that
using a separate communicator per thread would allow the user to
expose parallelism, they do not compare MPI Endpoints with the
communicator-based approach. Our work compares the capabilities of the
existing MPI standard with user-visible endpoints, demonstrating
scenarios where VCIs do as well as endpoints and where they falter.
Additionally, their work does not describe the notion of progress,
which is critical for correctness. Our work, on the other hand, does
not sacrifice correctness for performance.

\subsection{MPI libraries}

\minititle{Open MPI}. A couple of
works~\cite{gopalakrishnan2019improved,patinyasakdikul2019give}
on Open MPI are conceptually similar to our work---they use
fine-grained critical sections and map parallelism
available in the existing MPI standard to multiple network
hardware contexts to improve \threads{} communication
performance. However, both works do not compare
against user-visible endpoints or MPI everywhere.
Additionally, like the MPI Endpoints work, neither of these works
discusses the notion of shared progress, ignoring correctness.

Gopalkrishnan et al.~\cite{gopalakrishnan2019improved} evaluate
the communication performance of \threads{} with OFI scalable endpoints. Recognizing the
practical performance limitations of scalable endpoints, our
work uses regular OFI endpoints instead (see \secref{sec:multi_vcis}),
and hence we observe much larger speedups than their work obtains.

Similar to VCIs in our work, Patinyasakdikul et al.~\cite{patinyasakdikul2019give}
define Communication Resource Instances (CRIs). Their approach
involves creating a pool of CRIs and either assigning CRIs to
operations in a round-robin fashion or assigning CRIs to threads
using thread-local storage. While this approach may be correct for a subset
of operations, some CRIs break MPI's semantics for operations such as
\mpiaccum{} operations to the same target location. Such operations
are ordered by default on a window. In terms of performance, even with user-exposed parallelism
their point-to-point communication performance does not
scale with increasing number of threads unlike the results of our work.    

\begin{sloppypar}
\minititle{Intel MPI}. Since its 2019 release, the Intel MPI
library has utilized multiple network hardware contexts on Intel
Omni-Path through its multiple endpoints support~\cite{intelmpi}.
However, this support is only for a nonstandard threading level:
MPI\_THREAD\_SPLIT, which does not cover all cases possible
in the MPI\_THREAD\_MULTIPLE threading level. In contrast, our work with
VCIs fully and correctly supports MPI\_THREAD\_MULTIPLE.
\end{sloppypar}

In this work, we do not compare against the capabilities of other MPI libraries
since our goal is not to show that we can do
better than they can; rather, our aim is to study the strengths and limitations of
MPI-3.1 compared with those of user-visible endpoints. Adding other
libraries into the mix would blur the analysis in this paper. However,
a separate performance comparison with other libraries would make
a worthy future study.

\subsection{Distributed-memory programming models}

The newest version of the OpenSHMEM
specification features user-visible network contexts.  Dinan et
al.~\cite{dinan2014contexts} evaluate this approach. Although our work
is not on OpenSHMEM, the motivation to improve multithreaded
communication is the same. Instead of leaping into extending the
standard, however, we evaluate the capabilities of the existing
MPI-3.1 standard for \threads. Given the strengths and limitations of
MPI showcased in this paper, the MPI community is better equipped to
propose extensions to MPI, if any.

\section{Concluding remarks}
\label{sec:conclusion}

The \threads{} programming model is critical for effectively utilizing
modern processors. To dissolve its communication bottleneck,
however, domain scientists must expose logical parallelism in their
communication. Only then will we able to achieve the true
potential of \threads{}. The school of thought so far has been that
we need user-visible endpoints to express logical parallelism.
In this paper, however, we show that the existing MPI standard
already allows its users to overcome its ordering constraints
and express parallelism. By mapping MPI-3.1's  parallelism 
to internal virtual communication interfaces, in the majority of cases
we can achieve communication performance equal to the performance of
user-visible endpoints and MPI everywhere without sacrificing
correctness. More important, domain scientists do not need to
worry about managing and mapping to the limited hardware
resources with MPI 3.1, which is not the case in a user-visible
solution such as MPI Endpoints. We expect that MPI-4.0 will
increase the opportunities for users to express parallelism
through hints, and that, with the adoption of VCIs, MPI developers will be
able to effectively exploit new and current ways of expressing
logical communication parallelism.
\begin{acks}
We gratefully acknowledge the computing resources provided
and operated by the Joint Laboratory for System Evaluation (JLSE)
at Argonne National Laboratory (ANL). We thank the continuous feedback
from the members of the PMRS group at ANL, and we thank Gail Pieper
from ANL for her timely edits on this paper. This
work is supported by the U.S. Department of Energy, Office of
Science, under contract DE-AC02-06CH11357.
\end{acks}

\bibliographystyle{ACM-Reference-Format}
\bibliography{bib/refsused}


\begin{thebibliography}{42}


\ifx \showCODEN    \undefined \def \showCODEN     #1{\unskip}     \fi
\ifx \showDOI      \undefined \def \showDOI       #1{#1}\fi
\ifx \showISBNx    \undefined \def \showISBNx     #1{\unskip}     \fi
\ifx \showISBNxiii \undefined \def \showISBNxiii  #1{\unskip}     \fi
\ifx \showISSN     \undefined \def \showISSN      #1{\unskip}     \fi
\ifx \showLCCN     \undefined \def \showLCCN      #1{\unskip}     \fi
\ifx \shownote     \undefined \def \shownote      #1{#1}          \fi
\ifx \showarticletitle \undefined \def \showarticletitle #1{#1}   \fi
\ifx \showURL      \undefined \def \showURL       {\relax}        \fi
\providecommand\bibfield[2]{#2}
\providecommand\bibinfo[2]{#2}
\providecommand\natexlab[1]{#1}
\providecommand\showeprint[2][]{arXiv:#2}

\bibitem[\protect\citeauthoryear{??}{rza}{[n.d.]a}]%
        {rzambre_bspmm}
 \bibinfo{year}{[n.d.]}\natexlab{a}.
\newblock \bibinfo{title}{{BSPMM mini-app}}.
\newblock
\newblock
\newblock
\shownote{\url{https://github.com/rzambre/bspmm}.}


\bibitem[\protect\citeauthoryear{??}{ebm}{[n.d.]}]%
        {ebmsmini}
 \bibinfo{year}{[n.d.]}\natexlab{}.
\newblock \bibinfo{title}{{EBMS} mini-app}.
\newblock
\newblock
\newblock
\shownote{\url{https://github.com/ANL-CESAR/EBMS}.}


\bibitem[\protect\citeauthoryear{??}{rza}{[n.d.]b}]%
        {rzambre_ebms}
 \bibinfo{year}{[n.d.]}\natexlab{b}.
\newblock \bibinfo{title}{{Extended EBMS mini-app}}.
\newblock
\newblock
\newblock
\shownote{\url{https://github.com/rzambre/ebms}.}


\bibitem[\protect\citeauthoryear{??}{hfi}{[n.d.]}]%
        {hfi_guide}
 \bibinfo{year}{[n.d.]}\natexlab{}.
\newblock \bibinfo{title}{{I}ntel {O}mni-{P}ath {F}abric {H}ost {S}oftware}.
\newblock
\newblock
\newblock
\shownote{\url{https://www.intel.com/content/dam/support/us/en/documents/network-and-i-o/fabric-products/Intel_OP_Fabric_Host_Software_UG_H76470_v9_0.pdf}.}


\bibitem[\protect\citeauthoryear{??}{int}{[n.d.]}]%
        {intelmpi}
 \bibinfo{year}{[n.d.]}\natexlab{}.
\newblock \bibinfo{title}{{Intel\textsuperscript{\textregistered} MPI Multiple
  Endpoints Support}}.
\newblock
\newblock
\newblock
\shownote{\url{https://software.intel.com/en-us/mpi-developer-guide-linux-multiple-endpoints-support}.}


\bibitem[\protect\citeauthoryear{??}{mlx}{[n.d.]}]%
        {mlxRPM}
 \bibinfo{year}{[n.d.]}\natexlab{}.
\newblock \bibinfo{title}{{M}ellanox {P}{R}{M}}.
\newblock
\newblock
\newblock
\shownote{\url{http://www.mellanox.com/related-docs/user_manuals/Ethernet_Adapters_Programming_Manual.pdf}.}


\bibitem[\protect\citeauthoryear{??}{mpi}{[n.d.]a}]%
        {mpi4draft}
 \bibinfo{year}{[n.d.]}\natexlab{a}.
\newblock \bibinfo{title}{{MPI-4.0 Draft Report}}.
\newblock
\newblock
\newblock
\shownote{\url{https://www.mpi-forum.org/docs/drafts/mpi-2019-draft-report.pdf}.}


\bibitem[\protect\citeauthoryear{??}{mpi}{[n.d.]b}]%
        {mpiendpointsproposal}
 \bibinfo{year}{[n.d.]}\natexlab{b}.
\newblock \bibinfo{title}{{MPI} {E}ndpoints}.
\newblock
\newblock
\newblock
\shownote{\url{https://github.com/mpi-forum/mpi-issues/issues/56}.}


\bibitem[\protect\citeauthoryear{??}{non}{[n.d.]}]%
        {nonovertakingorder}
 \bibinfo{year}{[n.d.]}\natexlab{}.
\newblock \bibinfo{title}{{S}emantics of {P}oint-to-{P}oint {C}ommunication}.
\newblock
\newblock
\newblock
\shownote{\url{https://www.mpi-forum.org/docs/mpi-3.1/mpi31-report/node58.htm}.}


\bibitem[\protect\citeauthoryear{??}{avs}{[n.d.]}]%
        {avsharing}
 \bibinfo{year}{[n.d.]}\natexlab{}.
\newblock \bibinfo{title}{Shared {AV} table in {OFI/PSM2}}.
\newblock
\newblock
\newblock
\shownote{\url{https://github.com/ofiwg/libfabric/issues/5080}.}


\bibitem[\protect\citeauthoryear{??}{top}{[n.d.]}]%
        {top500interconnects}
 \bibinfo{year}{[n.d.]}\natexlab{}.
\newblock \bibinfo{title}{{TOP}500 Meanderings: {I}nfini{B}and {F}ends {O}ff
  {S}upercomputing {C}hallengers}.
\newblock
\newblock
\newblock
\shownote{\url{https://www.top500.org/news/top500-meanderings-infiniband-fends-off-supercomputing-challengers/
  }.}


\bibitem[\protect\citeauthoryear{Amer, Archer, Blocksome, Cao, Chuvelev,
  Fujita, Garzaran, Guo, Hammond, Iwasaki, et~al\mbox{.}}{Amer
  et~al\mbox{.}}{2019}]%
        {amer2019software}
\bibfield{author}{\bibinfo{person}{Abdelhalim Amer}, \bibinfo{person}{Charles
  Archer}, \bibinfo{person}{Michael Blocksome}, \bibinfo{person}{Chongxiao
  Cao}, \bibinfo{person}{Michael Chuvelev}, \bibinfo{person}{Hajime Fujita},
  \bibinfo{person}{Maria Garzaran}, \bibinfo{person}{Yanfei Guo},
  \bibinfo{person}{Jeff~R Hammond}, \bibinfo{person}{Shintaro Iwasaki},
  {et~al\mbox{.}}} \bibinfo{year}{2019}\natexlab{}.
\newblock \showarticletitle{Software combining to mitigate multithreaded MPI
  contention}. In \bibinfo{booktitle}{\emph{Proceedings of the ACM
  International Conference on Supercomputing}}. ACM, \bibinfo{pages}{367--379}.
\newblock


\bibitem[\protect\citeauthoryear{Amer, Lu, Wei, Balaji, and Matsuoka}{Amer
  et~al\mbox{.}}{2015}]%
        {amer2015mpi+}
\bibfield{author}{\bibinfo{person}{Abdelhalim Amer}, \bibinfo{person}{Huiwei
  Lu}, \bibinfo{person}{Yanjie Wei}, \bibinfo{person}{Pavan Balaji}, {and}
  \bibinfo{person}{Satoshi Matsuoka}.} \bibinfo{year}{2015}\natexlab{}.
\newblock \showarticletitle{MPI+ threads: Runtime contention and remedies}.
\newblock \bibinfo{journal}{\emph{ACM SIGPLAN Notices}} \bibinfo{volume}{50},
  \bibinfo{number}{8} (\bibinfo{year}{2015}), \bibinfo{pages}{239--248}.
\newblock


\bibitem[\protect\citeauthoryear{Balaji, Buntinas, Goodell, Gropp, Krishna,
  Lusk, and Thakur}{Balaji et~al\mbox{.}}{2010b}]%
        {balaji2010pmi}
\bibfield{author}{\bibinfo{person}{Pavan Balaji}, \bibinfo{person}{Darius
  Buntinas}, \bibinfo{person}{David Goodell}, \bibinfo{person}{William Gropp},
  \bibinfo{person}{Jayesh Krishna}, \bibinfo{person}{Ewing Lusk}, {and}
  \bibinfo{person}{Rajeev Thakur}.} \bibinfo{year}{2010}\natexlab{b}.
\newblock \showarticletitle{PMI: A scalable parallel process-management
  interface for extreme-scale systems}. In \bibinfo{booktitle}{\emph{European
  MPI Users' Group Meeting}}. Springer, \bibinfo{pages}{31--41}.
\newblock


\bibitem[\protect\citeauthoryear{Balaji, Buntinas, Goodell, Gropp, and
  Thakur}{Balaji et~al\mbox{.}}{2008}]%
        {balaji2008toward}
\bibfield{author}{\bibinfo{person}{Pavan Balaji}, \bibinfo{person}{Darius
  Buntinas}, \bibinfo{person}{David Goodell}, \bibinfo{person}{William Gropp},
  {and} \bibinfo{person}{Rajeev Thakur}.} \bibinfo{year}{2008}\natexlab{}.
\newblock \showarticletitle{Toward efficient support for multithreaded MPI
  communication}. In \bibinfo{booktitle}{\emph{European Parallel Virtual
  Machine/Message Passing Interface Users’ Group Meeting}}. Springer,
  \bibinfo{pages}{120--129}.
\newblock


\bibitem[\protect\citeauthoryear{Balaji, Buntinas, Goodell, Gropp, and
  Thakur}{Balaji et~al\mbox{.}}{2010a}]%
        {balaji2010fine}
\bibfield{author}{\bibinfo{person}{Pavan Balaji}, \bibinfo{person}{Darius
  Buntinas}, \bibinfo{person}{David Goodell}, \bibinfo{person}{William Gropp},
  {and} \bibinfo{person}{Rajeev Thakur}.} \bibinfo{year}{2010}\natexlab{a}.
\newblock \showarticletitle{Fine-grained multithreading support for hybrid
  threaded MPI programming}.
\newblock \bibinfo{journal}{\emph{The International Journal of High Performance
  Computing Applications}} \bibinfo{volume}{24}, \bibinfo{number}{1}
  (\bibinfo{year}{2010}), \bibinfo{pages}{49--57}.
\newblock


\bibitem[\protect\citeauthoryear{Bauer, Treichler, Slaughter, and Aiken}{Bauer
  et~al\mbox{.}}{2012}]%
        {bauer2012legion}
\bibfield{author}{\bibinfo{person}{Michael Bauer}, \bibinfo{person}{Sean
  Treichler}, \bibinfo{person}{Elliott Slaughter}, {and} \bibinfo{person}{Alex
  Aiken}.} \bibinfo{year}{2012}\natexlab{}.
\newblock \showarticletitle{Legion: Expressing locality and independence with
  logical regions}. In \bibinfo{booktitle}{\emph{SC'12: Proceedings of the
  International Conference on High Performance Computing, Networking, Storage
  and Analysis}}. IEEE, \bibinfo{pages}{1--11}.
\newblock


\bibitem[\protect\citeauthoryear{Bernholdt, Boehm, Bosilca, Gorentla~Venkata,
  Grant, Naughton, Pritchard, Schulz, and Vallee}{Bernholdt
  et~al\mbox{.}}{2017}]%
        {bernholdt2017survey}
\bibfield{author}{\bibinfo{person}{David~E Bernholdt}, \bibinfo{person}{Swen
  Boehm}, \bibinfo{person}{George Bosilca}, \bibinfo{person}{Manjunath
  Gorentla~Venkata}, \bibinfo{person}{Ryan~E Grant}, \bibinfo{person}{Thomas
  Naughton}, \bibinfo{person}{Howard~P Pritchard}, \bibinfo{person}{Martin
  Schulz}, {and} \bibinfo{person}{Geoffroy~R Vallee}.}
  \bibinfo{year}{2017}\natexlab{}.
\newblock \showarticletitle{A survey of MPI usage in the U.S. Exascale
  Computing Project}.
\newblock \bibinfo{journal}{\emph{Concurrency and Computation: Practice and
  Experience}} (\bibinfo{year}{2017}), \bibinfo{pages}{e4851}.
\newblock


\bibitem[\protect\citeauthoryear{Bulu{\c{c}}, Beamer, Madduri, Asanovic, and
  Patterson}{Bulu{\c{c}} et~al\mbox{.}}{2017}]%
        {bulucc2017distributed}
\bibfield{author}{\bibinfo{person}{Aydin Bulu{\c{c}}}, \bibinfo{person}{Scott
  Beamer}, \bibinfo{person}{Kamesh Madduri}, \bibinfo{person}{Krste Asanovic},
  {and} \bibinfo{person}{David Patterson}.} \bibinfo{year}{2017}\natexlab{}.
\newblock \showarticletitle{Distributed-memory breadth-first search on massive
  graphs}.
\newblock \bibinfo{journal}{\emph{arXiv preprint arXiv:1705.04590}}
  (\bibinfo{year}{2017}).
\newblock


\bibitem[\protect\citeauthoryear{Dinan, Balaji, Goodell, Miller, Snir, and
  Thakur}{Dinan et~al\mbox{.}}{2013}]%
        {dinan2013enabling}
\bibfield{author}{\bibinfo{person}{James Dinan}, \bibinfo{person}{Pavan
  Balaji}, \bibinfo{person}{David Goodell}, \bibinfo{person}{Douglas Miller},
  \bibinfo{person}{Marc Snir}, {and} \bibinfo{person}{Rajeev Thakur}.}
  \bibinfo{year}{2013}\natexlab{}.
\newblock \showarticletitle{Enabling MPI interoperability through flexible
  communication endpoints}. In \bibinfo{booktitle}{\emph{Proceedings of the
  20th European MPI Users' Group Meeting}}. ACM, \bibinfo{pages}{13--18}.
\newblock


\bibitem[\protect\citeauthoryear{Dinan and Flajslik}{Dinan and
  Flajslik}{2014}]%
        {dinan2014contexts}
\bibfield{author}{\bibinfo{person}{James Dinan} {and} \bibinfo{person}{Mario
  Flajslik}.} \bibinfo{year}{2014}\natexlab{}.
\newblock \showarticletitle{Contexts: a mechanism for high throughput
  communication in OpenSHMEM}. In \bibinfo{booktitle}{\emph{Proceedings of the
  8th International Conference on Partitioned Global Address Space Programming
  Models}}. ACM, \bibinfo{pages}{10}.
\newblock


\bibitem[\protect\citeauthoryear{Dinan, Grant, Balaji, Goodell, Miller, Snir,
  and Thakur}{Dinan et~al\mbox{.}}{2014}]%
        {dinan2014enabling}
\bibfield{author}{\bibinfo{person}{James Dinan}, \bibinfo{person}{Ryan~E
  Grant}, \bibinfo{person}{Pavan Balaji}, \bibinfo{person}{David Goodell},
  \bibinfo{person}{Douglas Miller}, \bibinfo{person}{Marc Snir}, {and}
  \bibinfo{person}{Rajeev Thakur}.} \bibinfo{year}{2014}\natexlab{}.
\newblock \showarticletitle{Enabling communication concurrency through flexible
  {MPI} endpoints}.
\newblock \bibinfo{journal}{\emph{The International Journal of HPC
  Applications}} \bibinfo{volume}{28}, \bibinfo{number}{4}
  (\bibinfo{year}{2014}), \bibinfo{pages}{390--405}.
\newblock


\bibitem[\protect\citeauthoryear{Felker, Siegel, Smith, Romano, and
  Forget}{Felker et~al\mbox{.}}{2014}]%
        {felker2014energy}
\bibfield{author}{\bibinfo{person}{Kyle~G Felker}, \bibinfo{person}{Andrew~R
  Siegel}, \bibinfo{person}{Kord~S Smith}, \bibinfo{person}{Paul~K Romano},
  {and} \bibinfo{person}{Benoit Forget}.} \bibinfo{year}{2014}\natexlab{}.
\newblock \showarticletitle{The energy band memory server algorithm for
  parallel Monte Carlo transport calculations}. In
  \bibinfo{booktitle}{\emph{SNA+ MC 2013-Joint International Conference on
  Supercomputing in Nuclear Applications+ Monte Carlo}}. EDP Sciences,
  \bibinfo{pages}{04207}.
\newblock


\bibitem[\protect\citeauthoryear{Gopalakrishnan, Cabral, Erwin, and
  Ganapathi}{Gopalakrishnan et~al\mbox{.}}{2019}]%
        {gopalakrishnan2019improved}
\bibfield{author}{\bibinfo{person}{Aravind Gopalakrishnan},
  \bibinfo{person}{Matias~A Cabral}, \bibinfo{person}{James~P Erwin}, {and}
  \bibinfo{person}{Ravindra~Babu Ganapathi}.} \bibinfo{year}{2019}\natexlab{}.
\newblock \showarticletitle{Improved MPI Multi-Threaded Performance using OFI
  Scalable Endpoints}. In \bibinfo{booktitle}{\emph{2019 IEEE Symposium on
  High-Performance Interconnects (HOTI)}}. IEEE, \bibinfo{pages}{36--39}.
\newblock


\bibitem[\protect\citeauthoryear{Grant, Dosanjh, Levenhagen, Brightwell, and
  Skjellum}{Grant et~al\mbox{.}}{2019}]%
        {grant2019finepoints}
\bibfield{author}{\bibinfo{person}{Ryan~E Grant}, \bibinfo{person}{Matthew~GF
  Dosanjh}, \bibinfo{person}{Michael~J Levenhagen}, \bibinfo{person}{Ron
  Brightwell}, {and} \bibinfo{person}{Anthony Skjellum}.}
  \bibinfo{year}{2019}\natexlab{}.
\newblock \showarticletitle{Finepoints: Partitioned multithreaded mpi
  communication}. In \bibinfo{booktitle}{\emph{International Conference on High
  Performance Computing}}. Springer, \bibinfo{pages}{330--350}.
\newblock


\bibitem[\protect\citeauthoryear{Grun, Hefty, Sur, Goodell, Russell, Pritchard,
  and Squyres}{Grun et~al\mbox{.}}{2015}]%
        {grun2015brief}
\bibfield{author}{\bibinfo{person}{Paul Grun}, \bibinfo{person}{Sean Hefty},
  \bibinfo{person}{Sayantan Sur}, \bibinfo{person}{David Goodell},
  \bibinfo{person}{Robert~D Russell}, \bibinfo{person}{Howard Pritchard}, {and}
  \bibinfo{person}{Jeffrey~M Squyres}.} \bibinfo{year}{2015}\natexlab{}.
\newblock \showarticletitle{A brief introduction to the OpenFabrics
  Interfaces--a new network API for maximizing high performance application
  efficiency}. In \bibinfo{booktitle}{\emph{2015 IEEE 23rd Annual Symposium on
  High-Performance Interconnects}}. IEEE, \bibinfo{pages}{34--39}.
\newblock


\bibitem[\protect\citeauthoryear{Higgins, Probert, Hasnip, Refson, and
  Bush}{Higgins et~al\mbox{.}}{2015}]%
        {higgins2015hybrid}
\bibfield{author}{\bibinfo{person}{Edward Higgins}, \bibinfo{person}{Matt
  Probert}, \bibinfo{person}{Phil Hasnip}, \bibinfo{person}{Keith Refson},
  {and} \bibinfo{person}{Ian Bush}.} \bibinfo{year}{2015}\natexlab{}.
\newblock \bibinfo{booktitle}{\emph{Hybrid OpenMP and MPI within the CASTEP
  code}}.
\newblock \bibinfo{type}{{T}echnical {R}eport}. \bibinfo{institution}{ARCHER
  eCSE Technical Report}.
\newblock


\bibitem[\protect\citeauthoryear{Hoefler, Dinan, Buntinas, Balaji, Barrett,
  Brightwell, Gropp, Kale, and Thakur}{Hoefler et~al\mbox{.}}{2013}]%
        {hoefler2013mpi+}
\bibfield{author}{\bibinfo{person}{Torsten Hoefler}, \bibinfo{person}{James
  Dinan}, \bibinfo{person}{Darius Buntinas}, \bibinfo{person}{Pavan Balaji},
  \bibinfo{person}{Brian Barrett}, \bibinfo{person}{Ron Brightwell},
  \bibinfo{person}{William Gropp}, \bibinfo{person}{Vivek Kale}, {and}
  \bibinfo{person}{Rajeev Thakur}.} \bibinfo{year}{2013}\natexlab{}.
\newblock \showarticletitle{{MPI}+ {MPI}: A new hybrid approach to parallel
  programming with {MPI} plus shared memory}.
\newblock \bibinfo{journal}{\emph{Computing}} \bibinfo{volume}{95},
  \bibinfo{number}{12} (\bibinfo{year}{2013}), \bibinfo{pages}{1121--1136}.
\newblock


\bibitem[\protect\citeauthoryear{Holmes}{Holmes}{[n.d.]}]%
        {holmesintroducing}
\bibfield{author}{\bibinfo{person}{Daniel Holmes}.}
  \bibinfo{year}{[n.d.]}\natexlab{}.
\newblock \showarticletitle{Introducing Endpoints into the EMPI4Re MPI
  library}.
\newblock  (\bibinfo{year}{[n.\,d.]}).
\newblock


\bibitem[\protect\citeauthoryear{Jin, Jespersen, Mehrotra, Biswas, Huang, and
  Chapman}{Jin et~al\mbox{.}}{2011}]%
        {jin2011high}
\bibfield{author}{\bibinfo{person}{Haoqiang Jin}, \bibinfo{person}{Dennis
  Jespersen}, \bibinfo{person}{Piyush Mehrotra}, \bibinfo{person}{Rupak
  Biswas}, \bibinfo{person}{Lei Huang}, {and} \bibinfo{person}{Barbara
  Chapman}.} \bibinfo{year}{2011}\natexlab{}.
\newblock \showarticletitle{High performance computing using MPI and OpenMP on
  multi-core parallel systems}.
\newblock \bibinfo{journal}{\emph{Parallel Comput.}} \bibinfo{volume}{37},
  \bibinfo{number}{9} (\bibinfo{year}{2011}), \bibinfo{pages}{562--575}.
\newblock


\bibitem[\protect\citeauthoryear{Nieplocha, Harrison, and
  Littlefield}{Nieplocha et~al\mbox{.}}{1994}]%
        {nieplocha1994global}
\bibfield{author}{\bibinfo{person}{Jaroslaw Nieplocha},
  \bibinfo{person}{Robert~J Harrison}, {and} \bibinfo{person}{Richard~J
  Littlefield}.} \bibinfo{year}{1994}\natexlab{}.
\newblock \showarticletitle{Global arrays: a portable shared-memory programming
  model for distributed memory computers}. In
  \bibinfo{booktitle}{\emph{Proceedings of the 1994 ACM/IEEE conference on
  Supercomputing}}. IEEE Computer Society Press, \bibinfo{pages}{340--349}.
\newblock


\bibitem[\protect\citeauthoryear{Patinyasakdikul, Eberius, Bosilca, and
  Hjelm}{Patinyasakdikul et~al\mbox{.}}{2019}]%
        {patinyasakdikul2019give}
\bibfield{author}{\bibinfo{person}{Thananon Patinyasakdikul},
  \bibinfo{person}{David Eberius}, \bibinfo{person}{George Bosilca}, {and}
  \bibinfo{person}{Nathan Hjelm}.} \bibinfo{year}{2019}\natexlab{}.
\newblock \showarticletitle{Give MPI Threading a Fair Chance: A Study of
  Multithreaded MPI Designs}. In \bibinfo{booktitle}{\emph{2019 IEEE
  International Conference on Cluster Computing (CLUSTER)}}. IEEE.
\newblock


\bibitem[\protect\citeauthoryear{Paul F.~Fischer and Kerkemeier}{Paul
  F.~Fischer and Kerkemeier}{2008}]%
        {nek5000-web-page}
\bibfield{author}{\bibinfo{person}{James W.~Lottes Paul F.~Fischer} {and}
  \bibinfo{person}{Stefan~G. Kerkemeier}.} \bibinfo{year}{2008}\natexlab{}.
\newblock \bibinfo{title}{{nek5000} {W}eb page}.
\newblock
\newblock
\newblock
\shownote{http://nek5000.mcs.anl.gov.}


\bibitem[\protect\citeauthoryear{Plimpton}{Plimpton}{1995}]%
        {plimpton1995fast}
\bibfield{author}{\bibinfo{person}{Steve Plimpton}.}
  \bibinfo{year}{1995}\natexlab{}.
\newblock \showarticletitle{Fast parallel algorithms for short-range molecular
  dynamics}.
\newblock \bibinfo{journal}{\emph{J. Comput. Phys.}} \bibinfo{volume}{117},
  \bibinfo{number}{1} (\bibinfo{year}{1995}), \bibinfo{pages}{1--19}.
\newblock


\bibitem[\protect\citeauthoryear{Rabenseifner, Hager, and Jost}{Rabenseifner
  et~al\mbox{.}}{2009}]%
        {rabenseifner2009hybrid}
\bibfield{author}{\bibinfo{person}{Rolf Rabenseifner}, \bibinfo{person}{Georg
  Hager}, {and} \bibinfo{person}{Gabriele Jost}.}
  \bibinfo{year}{2009}\natexlab{}.
\newblock \showarticletitle{Hybrid MPI/OpenMP parallel programming on clusters
  of multi-core SMP nodes}. In \bibinfo{booktitle}{\emph{2009 17th Euromicro
  International Conference on Parallel, Distributed and Network-based
  Processing}}. IEEE, \bibinfo{pages}{427--436}.
\newblock


\bibitem[\protect\citeauthoryear{Raffenetti, Amer, Oden, Archer, Bland, Fujita,
  Guo, Janjusic, Durnov, Blocksome, et~al\mbox{.}}{Raffenetti
  et~al\mbox{.}}{2017}]%
        {raffenetti2017mpi}
\bibfield{author}{\bibinfo{person}{Ken Raffenetti}, \bibinfo{person}{Abdelhalim
  Amer}, \bibinfo{person}{Lena Oden}, \bibinfo{person}{Charles Archer},
  \bibinfo{person}{Wesley Bland}, \bibinfo{person}{Hajime Fujita},
  \bibinfo{person}{Yanfei Guo}, \bibinfo{person}{Tomislav Janjusic},
  \bibinfo{person}{Dmitry Durnov}, \bibinfo{person}{Michael Blocksome},
  {et~al\mbox{.}}} \bibinfo{year}{2017}\natexlab{}.
\newblock \showarticletitle{Why is MPI so slow?: Analyzing the fundamental
  limits in implementing MPI-3.1}. In \bibinfo{booktitle}{\emph{Proceedings of
  the International Conference for High Performance Computing, Networking,
  Storage and Analysis}}. ACM, \bibinfo{pages}{62}.
\newblock


\bibitem[\protect\citeauthoryear{Romano, Horelik, Herman, Nelson, Forget, and
  Smith}{Romano et~al\mbox{.}}{2014}]%
        {romano2014openmc}
\bibfield{author}{\bibinfo{person}{Paul~K Romano}, \bibinfo{person}{Nicholas~E
  Horelik}, \bibinfo{person}{Bryan~R Herman}, \bibinfo{person}{Adam~G Nelson},
  \bibinfo{person}{Benoit Forget}, {and} \bibinfo{person}{Kord Smith}.}
  \bibinfo{year}{2014}\natexlab{}.
\newblock \showarticletitle{OpenMC: A state-of-the-art Monte Carlo code for
  research and development}. In \bibinfo{booktitle}{\emph{SNA+ MC 2013-Joint
  International Conference on Supercomputing in Nuclear Applications+ Monte
  Carlo}}. EDP Sciences, \bibinfo{pages}{06016}.
\newblock


\bibitem[\protect\citeauthoryear{Shamis et~al\mbox{.}}{Shamis
  et~al\mbox{.}}{2015}]%
        {shamis2015ucx}
\bibfield{author}{\bibinfo{person}{Pavel Shamis} {et~al\mbox{.}}}
  \bibinfo{year}{2015}\natexlab{}.
\newblock \showarticletitle{UCX: an open source framework for HPC network APIs
  and beyond}. In \bibinfo{booktitle}{\emph{2015 IEEE 23rd Annual Symposium on
  High-Performane Interconnects.}} IEEE, \bibinfo{pages}{40--43}.
\newblock


\bibitem[\protect\citeauthoryear{Siegel, Smith, Felker, Romano, Forget, and
  Beckman}{Siegel et~al\mbox{.}}{2014}]%
        {siegel2014improved}
\bibfield{author}{\bibinfo{person}{A Siegel}, \bibinfo{person}{Kord Smith},
  \bibinfo{person}{K Felker}, \bibinfo{person}{P Romano},
  \bibinfo{person}{Benoit Forget}, {and} \bibinfo{person}{P Beckman}.}
  \bibinfo{year}{2014}\natexlab{}.
\newblock \showarticletitle{Improved cache performance in Monte Carlo transport
  calculations using energy banding}.
\newblock \bibinfo{journal}{\emph{Computer Physics Communications}}
  \bibinfo{volume}{185}, \bibinfo{number}{4} (\bibinfo{year}{2014}),
  \bibinfo{pages}{1195--1199}.
\newblock


\bibitem[\protect\citeauthoryear{Sridharan, Dinan, and Kalamkar}{Sridharan
  et~al\mbox{.}}{2014}]%
        {sridharan2014enabling}
\bibfield{author}{\bibinfo{person}{Srinivas Sridharan}, \bibinfo{person}{James
  Dinan}, {and} \bibinfo{person}{Dhiraj~D Kalamkar}.}
  \bibinfo{year}{2014}\natexlab{}.
\newblock \showarticletitle{Enabling efficient multithreaded MPI communication
  through a library-based implementation of MPI endpoints}. In
  \bibinfo{booktitle}{\emph{Proceedings of the International Conference for
  High Performance Computing, Networking, Storage and Analysis}}. IEEE Press,
  \bibinfo{pages}{487--498}.
\newblock


\bibitem[\protect\citeauthoryear{Valiev, Bylaska, Govind, Kowalski, Straatsma,
  Dam, Wang, Nieplocha, Apra, Windus, et~al\mbox{.}}{Valiev
  et~al\mbox{.}}{2010}]%
        {valiev2010nwchem}
\bibfield{author}{\bibinfo{person}{Marat Valiev}, \bibinfo{person}{Eric~J
  Bylaska}, \bibinfo{person}{Niranjan Govind}, \bibinfo{person}{Karol
  Kowalski}, \bibinfo{person}{Tjerk~P Straatsma}, \bibinfo{person}{Hubertus
  JJ~Van Dam}, \bibinfo{person}{Dunyou Wang}, \bibinfo{person}{Jarek
  Nieplocha}, \bibinfo{person}{Edoardo Apra}, \bibinfo{person}{Theresa~L
  Windus}, {et~al\mbox{.}}} \bibinfo{year}{2010}\natexlab{}.
\newblock \showarticletitle{{NWC}hem: A comprehensive and scalable open-source
  solution for large scale molecular simulations}.
\newblock \bibinfo{journal}{\emph{Computer Physics Comm.}}
  \bibinfo{volume}{181}, \bibinfo{number}{9} (\bibinfo{year}{2010}),
  \bibinfo{pages}{1477--1489}.
\newblock


\bibitem[\protect\citeauthoryear{Zambre, Chandramowlishwaran, and
  Balaji}{Zambre et~al\mbox{.}}{2018}]%
        {zambre2018scalable}
\bibfield{author}{\bibinfo{person}{Rohit Zambre}, \bibinfo{person}{Aparna
  Chandramowlishwaran}, {and} \bibinfo{person}{Pavan Balaji}.}
  \bibinfo{year}{2018}\natexlab{}.
\newblock \showarticletitle{Scalable communication endpoints for MPI+ Threads
  applications}. In \bibinfo{booktitle}{\emph{2018 IEEE 24th International
  Conference on Parallel and Distributed Systems (ICPADS)}}. IEEE,
  \bibinfo{pages}{803--812}.
\newblock


\end{thebibliography}


\end{document}